\documentclass[10pt]{article}
\usepackage[colorlinks,linkcolor=black,citecolor=black]{hyperref}   

\hypersetup{colorlinks=true, linkcolor=purple, urlcolor=blue, citecolor=cyan, anchorcolor=black}

\usepackage[utf8]{inputenc}	
\usepackage[T1]{fontenc}
\usepackage{lineno}					
\usepackage{tikz} 					

\usepackage{url}            
\usepackage{booktabs}       
\usepackage{nicefrac}       
\usepackage{microtype}      
\usepackage{fancyhdr}       
\usepackage{xcolor,paralist,etoolbox}
\usepackage{color}
\usepackage{sidecap}
\usepackage{booktabs}
\usepackage{multirow}
\usepackage{multicol}
\usepackage{array}
\usepackage[outdir=./]{epstopdf}
\usepackage{pdflscape}
\usepackage{enumitem}
\usepackage{authblk}
\usepackage{cite}%
\usepackage{mciteplus}
\usepackage{bm}
\usepackage{amssymb}
\usepackage{amsmath}
\usepackage{amsthm}
\usepackage{eucal}
\usepackage{wasysym}
\usepackage{braket,mleftright}
\mleftright
\usepackage[version=3]{mhchem}
\usepackage{siunitx}
\usepackage{cancel}

\usepackage[euler]{textgreek}
\usepackage{listings}
\usepackage{listingsutf8}
\usepackage[ruled,linesnumbered]{algorithm2e}

\usepackage{newfloat}
\DeclareFloatingEnvironment[name={Supplementary Figure}]{suppfigure}
\usepackage{sidecap}
\sidecaptionvpos{figure}{c}

\definecolor{lime}{HTML}{A6CE39}

\newcommand*\neper{\mathrm{e}}
\newcommand*\diff{\mathrm{d}}

\title{The Steepest Slope toward a Quantum Few-body Solution:\\
Gradient Variational Methods for the Quantum\\ Few-body
Problem}

\usepackage{authblk}

\author[1]{\small Paolo Recchia}
\author[2]{\small Debabrota Basu}
\author[3]{\small Mario Gattobigio}
\author[3, 4]{\small Christian Miniatura}
\author[1, 5]{\small Stéphane Bressan}

\affil[1]{\footnotesize School of Computing, National University of Singapore, 117417, Singapore}
\affil[2]{\footnotesize Univ. Lille, Inria, CNRS, Centrale Lille, UMR 9189 - CRIStAL, F-59000 Lille, France}
\affil[3]{\footnotesize Universit\'e C\^ote d'Azur, CNRS, Institut de Physique de Nice, 17 rue julien Lauprêtre, 06200 Nice, France}
\affil[4]{\footnotesize Centre for Quantum Technologies, National University of Singapore, 117543, Singapore}
\affil[5]{\footnotesize International Research Laboratory on Artificial Intelligence, CNRS International Laboratory, 2955, Singapore}

\date{}

\begin{document}

\maketitle

\begin{abstract}
Quantum few-body systems are deceptively simple. Indeed, with the notable exception of a few special cases, their associated Schr\"{o}dinger equation cannot be solved analytically for more than two particles. One has to resort to approximation methods to tackle quantum few-body problems. In particular, variational methods have been proposed to ease numerical calculations and obtain precise solutions. One such method is the Stochastic Variational Method, which employs a stochastic search to determine the number and parameters of correlated Gaussian basis functions used to construct an ansatz of the wave function.  Stochastic methods, however, face numerical and optimization challenges as the number of particles increases. 

We introduce a family of gradient variational methods that replace stochastic search with gradient optimization. We comparatively and empirically evaluate the performance of the baseline Stochastic Variational Method, several instances of the gradient variational method family, and some hybrid methods for selected few-body problems. We show that gradient and hybrid methods can be more efficient and effective than the Stochastic Variational Method. We discuss the role of singularities, oscillations, and gradient optimization strategies in the performance of the respective methods.
\end{abstract}

\vspace{0.5cm}


\vspace{0.5cm}


\section{Introduction}
\label{sec:introduction}

The instances of the Schr\"{o}dinger equation that govern the dynamics of quantum few-body systems typically lack analytical solutions when more than two particles are involved, even when the interaction potentials are known. Physicists have devised numerical approximation methods to address quantum few-body problems. Among these methods, the most prevalent are variational methods~\cite{cohen2019quantum}, which leverage the Variational Principle~\cite{griffiths2018introduction}, also referred to as Ritz theorem~\cite{suzuki1998stochastic}. The Variational Principle states that the expectation value of the Hamiltonian for any state of a quantum system is greater than or equal to the ground state energy. The ground state energy is precisely the expectation value of the Hamiltonian in the ground state. This expectation value can be expressed as the Rayleigh quotient involving the Hamiltonian and the wave function of the state.

One state-of-the-art variational method is the Stochastic Variational Method~\cite{suzuki1998stochastic}. As its name suggests, this variational method employs a stochastic search to determine the parameters of correlated Gaussian basis functions used to construct an ansatz of the wave function. However, as the number of particles increases, the Stochastic Variational Method and its variants, such as stochastic variational methods using other families of basis functions, face numerical and optimization challenges. 

Yasuyuki Suzuki and K\'{a}lm\`{a}n Varga, discussing the stochastic search of the Stochastic Variational Method in their 1998 reference textbook~\cite{suzuki1998stochastic}, wrote ``\emph{A skeptic may say that one should, instead of the above gambling method, try a deterministic parameter search $[\cdots]$}''.
 
We propose revisiting this claim in light of modern gradient descent methods, like the Adam algorithm~\cite{kingma2014adam}, developed for training machine learning models. Specifically, we introduce a family of gradient variational methods that replace stochastic search with gradient optimization. We comparatively and empirically evaluate the performance of the baseline Stochastic Variational Method, several instances of the gradient variational method family, and some hybrid methods for selected few-body problems. We show that gradient and hybrid methods can be more efficient and effective than the state-of-the-art Stochastic Variational Method. We discuss the role of singularities, oscillations, and gradient optimization strategies in the performance of the respective methods.

This work primarily focuses on few-body systems composed of identical bosons. Under certain conditions, these systems exhibit universal properties, such as the intriguing Efimov effect~\cite{efimov1970energy}. Moreover, systems of identical bosons are effectively represented by Gaussian-shaped potentials~\cite{recchia2022gaussian, kievsky2021efimov}. Specifically, given that clusters of helium atoms serve as natural candidates for Efimov Physics~\cite{kolganova20114he}, we use these systems for testing purposes. 

The paper is organized as follows. In Section \ref{sec:SVM}, we delineate the core principles of the Stochastic Variational Method as initially introduced in references~\cite{kukulin1977stochastic, suzuki1998stochastic}. Moving on to Section~\ref{sec:rw}, and after a short review of the current literature about methods to tackle few-body systems, we illustrate alternative optimization techniques suitable for nonlinear parameters. Section~\ref{sec:meth} introduces our proposal to embed certain gradient-based optimization approaches into the framework of the Stochastic Variational Method. Lastly, in Section~\ref{sec:perfor}, we compare and illustrate the performance of these optimization schemes with the original Stochastic Variational Method. We then conclude and offer some perspectives.

\section{Background: the Stochastic Variational Method} 
\label{sec:SVM}

In this section, we outline the key features, advantages, and drawbacks of the state-of-the-art Stochastic Variational Method. Throughout the paper, we follow the notations and definitions provided in~\cite{suzuki1998stochastic} unless otherwise specified.

\subsection{The variational method for solving the Schr\"odinger equation}
\label{sec:sch_ritz}

Let us consider few-body systems of interacting particles with two- and three-body interactions. The Schr\"odinger equation governing such systems can be formulated as a linear algebra problem by introducing the linear and Hermitian Hamiltonian operator~\eqref{eq:hamiltonian}, where $\hbar$ is the Planck constant, $N$  is the number of particles,  $a$, $b$, and $c$ are the indices of the different particles, $m_a$ is the mass of a particle, $V$ and $W$ are the two-body and a three-body potentials, respectively, and $\bm r_a$ is the position of a particle
\begin{equation}
\label{eq:hamiltonian}
    H = \sum_{a=1}^N -\frac{\hbar^2}{2m_a} \frac{\partial^2}{\partial \bm{r_a}}  + \sum_{b > a = 1}^N V(\bm{r_a},\bm{r_b}) +  \sum_{c > b > a = 1}^N W(\bm{r_a},\bm{r_b}, \bm{r_c}), 
\end{equation}
The Hilbert space of the Hamiltonian is infinite-dimensional with both a continuous and discrete spectrum of eigenvalues. However, we focus on finding the bound states of the Hamiltonian, which allows us to consider the Schr\"odinger equation as the eigenvalue problem~\eqref{eq:sch_eigen} with a discrete spectrum of eigenvalues $E_n$ and $\Phi_n$ eigenfunctions, $E_1$ and $\Phi_1$ are the ground state energy and wavefunction respectively.
\begin{equation}
\label{eq:sch_eigen}
    H \Phi_n = E_n \Phi_n.
\end{equation}

The Ritz theorem~\cite{suzuki1998stochastic}, also called variational principle~\cite{cohen2019quantum, griffiths2018introduction}, states that, for an arbitrary function $\phi$ of the state space, the expectation value of $H$ in the state $\phi$ is larger than or equal to the ground state energy, as expressed in Equation~\eqref{eq:functional}. In addition, the theorem states that $E[\phi]$ equals $E_1$ if and only if $\phi$ is an eigenstate of $H$ with eigenvalue $E_1$
\begin{equation}
\label{eq:functional}
    \ E[\phi] = \frac{\braket{\phi | H| \phi}}{\braket{\phi|\phi}} \geq E_1.
\end{equation}

Ritz variational method~\cite{suzuki1998stochastic} leverages Ritz theorem. To find the lowest possible value approaching the ground state energy $E_1$, we need to minimize the functional in~\eqref{eq:functional} with respect to the trial function $\phi$, often called \emph{ansatz}. The minimization is achieved by imposing the stationary condition of~\eqref{eq:functional}. Typically, the ansatz takes the form of a sum over basis elements $\psi_k$ as expressed 
\begin{equation}
\label{eq:system_vecs}
\psi\approx \phi = \sum_{k=1}^{K} c_k \psi_k.
\end{equation}
With this ansatz and in conjunction with the stationary condition, the Schrödinger Equation~\eqref{eq:sch_eigen} transforms into a linear system of $K$ equations in terms of the coefficients $c_k$ (more details are in Appendix~\ref{app:solv_sec_equ}). This system can be solved as a generalized finite eigenvalue problem, often referred to as the secular equation
\begin{equation}
\label{eq:eigenvalue}
\mathcal{HC} = \mathcal{SC}E,
\end{equation}
where $\mathcal{H}{ij} = \braket{\psi_i | H | \psi_j}$ and $\mathcal{S}{ij} = \braket{\psi_i | \psi_j}$ represent the matrix elements of the Hamiltonian and overlap on the chosen basis, respectively. Here, $E = \text{diag}(E_i)_{i=1,\dots, K}$ represents the eigenvalues (in particular, the lowest $E_1$ is the ground state energy approximation we are interested in), and $\mathcal{C}$ denotes the matrix of eigenvectors with columns being the coefficients $c_k$ for each eigenvalue. 

Several important aspects of variational methods should be emphasized. Firstly, the approximation given by~\eqref{eq:system_vecs} reduces the eigenvalue problem~\eqref{eq:eigenvalue} to a finite-dimensional form. Another crucial theorem asserts that increasing the number of basis elements enhances the accuracy of computed eigenvalues and consequently the ground state energy~\cite{suzuki1998stochastic}. Secondly, various diagonalization techniques can be applied to solve the secular equation~\eqref{eq:eigenvalue}. In this study, we adopt the method detailed in~\cite{piela2006ideas}, as described in Appendix~\ref{app:solv_sec_equ}, simplifying the calculation by converting the generalized eigenvalue problem into a standard one. Thirdly, the effectiveness of variational methods heavily relies on the choice of the ansatz, particularly the family of basis functions denoted as ${\psi_k}$. 

For the sake of completeness, it is important to note that the generalized Ritz theorem empowers the variational methods for the approximation of the excited states of the system as well. Specifically, solving the secular equation \eqref{eq:eigenvalue} provides an estimate of the energy spectrum of the system under consideration \cite{suzuki1998stochastic}.

\subsection{Stochastic optimization and Gaussian basis in the Stochastic Variational Method}
\label{sec:bgSVM}

Like any other variational technique, the Stochastic Variational Method relies on the Ritz theorem. A key feature of the Stochastic Variational Method is the choice of the ansatz, which takes the form of a sum over a system of non-orthonormal Gaussian basis functions in configuration space that is parametrized by a finite number of continuous nonlinear $\alpha^{(k)}_{ab}$ parameters
\begin{equation}
\label{eq:gaussian_basis}
    \braket{\bm r_1, \dots, \bm r_N| \psi_k} = \mathrm{exp}\biggl[-\frac{1}{2} \sum_{b > a=1}^{N}  \bigg(\frac{\bm{r_a} - \bm{r_b}}{\alpha^{(k)}_{ab}}\bigg)^2 \biggr].
\end{equation}
We refer to this choice as the \emph{Gaussian basis}. To minimize the functional~\eqref{eq:functional}, we observe, based on~\eqref{eq:system_vecs} and~\eqref{eq:gaussian_basis}, that two types of parameters need to be determined: the set of linear variational parameters represented as $c_k$, and the set of nonlinear variational parameters denoted as $\alpha_{ab}^{(k)}$ (these are the coefficients in the Gaussian exponents). Determining the linear parameters $c_k$ is a straightforward task involving solving a linear problem, specifically the eigenvalue problem described in~\eqref{eq:eigenvalue}. On the other hand, finding the nonlinear parameters can be quite challenging, especially in cases where various local minima influence the parameter space. The Stochastic Variational Method adopts a stochastic search to address this challenge. Indeed, introducing random steps in the process allows for exploring the entire parameter space, including regions with local minima. The fundamental concept behind the original Stochastic Variational Method is to perform multiple trial evaluations at each step when a new basis element is added. Subsequently, only the trial returning the lowest eigenvalue is chosen~\cite{kukulin1977stochastic}. 

The effectiveness of the Gaussian basis in describing various types of few-body systems can be attributed to two main reasons. First, under given conditions, \emph{the Gaussian basis exhibits the property of being overcomplete}, meaning that removing a countable subset of functions does not affect the completeness of the system (more details about the requirements for completeness are discussed and proved in~\cite{klahn1977convergence, shaw2020completeness}). This characteristic enhances the method's robustness and reliability; in fact, rejecting a countable subset of basis elements is not a significant issue. Secondly, the Gaussian basis simplifies the computation of certain integrals arising during the evaluation of matrix elements in the eigenvalue problem~\eqref{eq:eigenvalue}, as a significant number of Gaussian integrals can be analytically evaluated. 

However, the overcompletness in the Gaussian basis introduces a significant drawback. When dealing with the generalized eigenvalue problem~\eqref{eq:eigenvalue}, the computation of the inverse of the overlap matrix $\mathcal{S}$ becomes necessary (see appendix~\ref{app:solv_sec_equ} for further details). Given that the basis vectors are overcomplete, some vectors may be linearly dependent on each other, resulting in very small eigenvalues of the overlap matrix $\mathcal{S}$. This, in turn, \textit{ contributes to numerical instabilities when attempting to calculate the inverse of} $\mathcal{S}$. In the Stochastic Variational Method, these singularities are eliminated whenever a newly proposed basis yields them. Subsequently, a fresh random basis is suggested. We emphasize that various stopping criteria can be suggested to stop the iterations in the Stochastic Variational Method. For example, the algorithm can conclude with a given number of basis $K$, or when a specified CPU runtime is reached or when the relative difference between the last two energy values attained is below a given threshold $\varepsilon$
\begin{equation}
\label{eq:stop_svm}
\bigg| \frac{E_k - E_{k-1}}{E_k} \bigg| \leq \varepsilon.
\end{equation}
In the next sections, as we mainly focus on a performance comparative analysis, we usually stop the Stochastic Variational Method after a given CPU runtime is reached. Another important aspect of the baseline Stochastic Variational Method pertains to the interval $(0, \alpha_{max}]$ within which the nonlinear variational parameters are randomly chosen. As depicted in Equation~\eqref{eq:gaussian_basis}, the variational parameters $\alpha^{(k)}_{ab}$ share the same dimensions as the interparticle distance. Consequently, we can deduce a reasonable range for selecting the nonlinear parameters based on a preliminary understanding of the system under consideration. Nevertheless, this information is often unavailable beforehand, presenting an additional challenge in implementing the Stochastic Variational Method.

The algorithm of the Stochastic Variational Method, as originally proposed in~\cite{kukulin1977stochastic}, is summarized in Algorithm~\ref{alg:or_SVM}, in Appendix~\ref{app:algs}.

\section{Related Work}
\label{sec:rw}

\subsection{Overview of methods for solving the quantum few-body problem}
\label{sec:overview_few_body}

Numerous methods have been developed to solve quantum few–body systems. The Stochastic Variational Method is one among them. Despite their variety, we can classify them into two big categories: \emph{non-variational methods} and \emph{variational methods}. These categories can be further divided into subcategories. Non-variational methods include the \emph{exact method} and \emph{numerical integration}. Variational methods can be further divided into \emph{variational methods with an orthonormal basis}, \emph{variational methods with a non-orthonormal basis}, and \emph{Variational Monte Carlo methods}.

The \emph{Faddeev Method}~\cite{faddeev1965mathematical} provides an exact technique for solving the Schr\"odinger equation in three-body systems. Initially developed for studying nuclear physics scattering problems, this method is also applicaple in identifying bound states within such systems. With specific classes of interactions, the Faddeev method solves three coupled single-particle Schr\"odinger equations, termed Faddeev components or equations, rather than the full Schr\"odinger equation. The appropriate boundary conditions applied to each Faddeev component yield the corresponding energy eigenvalue for the entire three-body system. Extensions of the Faddeev equations to systems beyond three-body are referred to as Faddeev-Yakubovsky methods~\cite{faddeev1993quantum}. However, scalability remains a primary challenge for this approach. The number of Faddeev components, and thus the differential equations to solve, grows factorially with the number of particles, making numerical computations quickly impractical. Presently, the Faddeev-Yakubovsky method has found applications in the literature for four-body systems~\cite{lazauskas2018solution, braaten2006universality}.

Among the \emph{numerical integration} techniques, the most common is the Numerov integration~\cite{numerov1927note}. It is a numerical method to solve ordinary differential equations of second order in which the first-order term does not appear, like the Schr\"odinger equation. Though it allows solving the Schr\"odinger equation with a wide variety of interactions, its application is restricted to low-dimensional cases: the current literature extends the Numerov integration in a grid of up four dimensions~\cite{kuenzer2016pushing, kuenzer2019four}.

\emph{Variational methods} require a choice of basis functions, which can be either orthonormal or non-orthonormal. \emph{Hyperspherical Harmonics} are a widely studied and applied example of orthonormal and complete basis functions. Although this basis was first introduced in 1935 by Zernike and Brinkman~\cite{zernike1935hyperspharische} and reintroduced 25 years later by Delves~\cite{delves1958tertiary} and Smith~\cite{smith1962symmetric}, it has only been in recent decades that it has seen widespread use and development~\cite{kievsky:2008_J.Phys.G}. The primary advantage of Hyperspherical Harmonics is their property of being eigenfunctions of the kinetic energy operator.
Additionally, because Hyperspherical Harmonics form an orthonormal basis, the overlap matrix $\mathcal{S}$ in~\eqref{eq:eigenvalue} becomes the identity matrix, simplifying the generalized eigenvalue problem in~\eqref{eq:eigenvalue} to a standard eigenvalue problem. This eliminates the need to calculate the inverse of the overlap matrix, thereby avoiding the singularity issues discussed in Section~\ref{sec:bgSVM}. However, unlike the Gaussian basis, the matrix elements of the potential term of the Hamiltonian with Hyperspherical Harmonics cannot be evaluated analytically, necessitating numerical integration techniques. For instance, calculations based on Chebyshev and Laguerre weights and points~\cite{abramowitz1964handbook} are used in~\cite{kievsky:2008_J.Phys.G}. Moreover, the number of Hyperspherical Harmonics required for convergence is typically larger than that of Gaussians~\cite{marcucci2020hyperspherical}. This large number of basis elements makes the diagonalization of the Hamiltonian matrix numerically challenging, and standard diagonalization techniques are often inadequate. In the Hyperspherical Harmonics method, the iterative Lanczos algorithm is widely used to address this issue~\cite{lanczos1950iteration, marcucci2020hyperspherical}.

In contrast to the variational methods with orthonormal basis, the other family of variational methods, such as the Stochastic Variational Method \cite{suzuki1998stochastic}, compute the nonlinear parameters of the chosen non-orthonormal basis via random trial and error. In literature, the Gaussian basis is considered state-of-the-art for the Stochastic Variational Method, and other bases have hardly been considered. Alternative families of bases, like exponentials and wavelets, are currently under investigation. For optimising the nonlinear parameters in those bases, the same arguments of the Stochastic Variational Method apply.    

The \emph{Variational Monte Carlo} method always relies on the variational principle, particularly focusing on the functional described by equation~\eqref{eq:functional}. By defining the local energy $ E_{\text{loc}}(\bm{r}) = \frac{H \phi(\bm{r})}{\phi(\bm{r})}$, we can express the functional~\eqref{eq:functional} as the average of $E_{\text{loc}}(\bm{r})$ where the normalized square of the ansatz $|\phi(\bm{r})|^2/\int \diff \bm{r} |\phi(\bm{r})|^2$ represents a probability distribution. We can then calculate the functional~\eqref{eq:functional} via Monte Carlo techniques where the points for the integration are sampled from the probability distribution $|\phi(\bm{r})|^2/\int \diff \bm{r} |\phi(\bm{r})|^2$. Various sampling techniques can be employed, such as Metropolis sampling~\cite{metropolis1953equation,hastings1970monte}, importance sampling through simulation of the Fokker-Planck differential equation~\cite{hammond1994monte}, and correlated sampling~\cite{bar1988mh}. Once the most suitable sampling technique is identified, the variational parameters of the ansatz $\phi$ are primarily optimized by minimizing the expectation value of the local energy. The stochasticity due to the sampling introduces a statistical uncertainty in the energy expectation value, which can be higher compared to other variational techniques. Although this uncertainty can be mitigated by increasing the sampling points, it may limit the method's performance.

\subsection{Nonlinear optimization in the Stochastic Variational Methods}
\label{sec:diff_optim}

Besides the singularities' issue introduced in Section \ref{sec:bgSVM}, another common issue of the Stochastic Variational Method concerns its computational cost, especially when the number $N$ of particles increases. Firstly, the number of non–linear parameters $\alpha_{ab}$ in the ansatz in Eqaution \eqref{eq:gaussian_basis} increases quadratically with $N$, like $N(N - 1)/2$. Secondly, to efficiently and effectively converge as the space of parameters increases, a larger number $K$ of basis functions is also required. Thirdly, to solve the secular equation~\eqref{eq:eigenvalue}, one must calculate the determinant and the inverse of a $K \times K$ matrix, which scales like O($K^3$) (Gauss-Jordan). Lastly, wavefunction symmetrization is required for systems of identical Bosons and Fermions, and $N!$ of the previously mentioned operations is necessary. The main bottleneck regards the nonlinear optimization of the variational parameters,  which is hard to accomplish, especially when the space is affected by multiple local minima. Nonlinear optimization is a broad field~\cite{aragon2019nonlinear}, and a comprehensive review of all existing approaches is beyond the scope of this paper. In this brief section, we offer an overview of two major categories: \emph{gradient-free optimization} and \emph{gradient-based optimization}.

\subsubsection{Gradient-free optimization methods}

Gradient-free optimization methods do not require information about the gradient of the objective function. They can be useful when this information is unavailable or the functions ae not smooth. We can classify them into three subcategories.

\emph{Deterministic derivative-free optimizations}, such as Nelder-Mead~\cite{nelder1965simplex}, Powell~\cite{powell1964efficient}, that move on the slope of the function following a given strategy. These approaches have different drawbacks. First, their high computational cost makes them not scalable for systems with increasing particles. Second, they are sensitive to a given starting point, always reaching the same minimum from the same initial condition. Third, the solutions found by these approaches may not be the global ones since they can easily get stuck in local minima, especially when the size of the Hilbert space increases.

\emph{Metaheuristic approaches}, for example, Genetic Algorithm~\cite{fogel1998evolutionary} and Simulated Annealing~\cite{kirkpatrick1983optimization}, are other advanced derivative-free techniques to find the global optimum of a given function. Even though they have never been appropriately tested for optimizing the nonlinear variational parameters, both these methods were born to handle tasks on a discrete space~\cite{bianchi2009survey}. In addition, both suffer from the local minima issues as the deterministic approaches. 

While the abovementioned optimization methods may seem more sophisticated than the \emph{stochastic optimization}, state-of-the-art research indicates that random trials often yield greater success by circumventing the common occurrence of getting stuck in local minima~\cite{suzuki1998stochastic}.

\subsubsection{Gradient-based optimization methods}
\label{sec:gd_optim}

In the current literature, no systematic benchmarking of \emph{gradient-based optimizations} applied to variational methods for quantum few-body problems exists. However, in past years, gradient-based optimization methods have become widespread due to the increasing popularity of machine learning models. When the objective function to optimize is smooth, the gradient-based methods proved to be reliable techniques for efficiently finding the minima of the function thanks to the calculation of its gradient. Gradient Descent~\cite{curry1944method} is the foundational optimization algorithm. It updates the parameters in the opposite direction of the gradient of the objective function. Some variants of gradient descent help speed the algorithm's convergence in different contexts. Gradient descent with momentum~\cite{rumelhart1986learning} stores the past gradients at each iteration and determines the next update as a linear combination of these past gradients. Gradient descent with Root Mean Square Propagation (RMSProp)~\cite{hinton2012neural} adjusts the learning rate by dividing it by a factor that depends on the moving average of the past squared gradients. Gradient descent with Adaptive Momentum Estimation (Adam)~\cite{kingma2014adam} combine the idea in Momentum and RMSProp. It adapts the learning rates by employing an exponential forgetting factor to compute running averages for both gradients and the squared of gradients. This latter gradient optimization method, in particular, has demonstrated superior performance compared to previous ones across different datasets and various machine learning models, as indicated in~\cite{kingma2014adam} (for a more detailed outline about the gradient methods, refer to~\cite{ruder2016overview}). The current paper aims to evaluate and compare the performance of gradient-based optimization techniques in the framework of variational methods applied to quantum few-body problems. Following initial testing, we have identified Adam as the most efficient gradient-based optimization. This finding suggests that Adam could serve as a promising alternative for exploring the nonlinear parameters in the Gaussian basis of the Stochastic Variational Method. 

It is crucial to recognize that during the approach to a local or global minimum, gradient-based optimizations can lead to oscillations in the loss function (representing the system energy in this context). Such oscillations are well-known across various gradient methods, including Adam~\cite{goodfellow2016deep, ruder2016overview}. Recent research has delved deeper into their underlying causes. For instance, in studies such as those~\cite{choo2020,arora2022understanding,ahn2022understanding}, the authors have investigated various neural network models, loss functions, and datasets. They have highlighted the relationship between the shape of the loss function near a local minimum and the selection of the learning rate as significant factors contributing to oscillations and potential divergences. Specifically, when the product of the learning rate and the sharpness of the loss function (the largest eigenvalue of its Hessian) exceeds two, gradient optimization becomes unstable. This instability manifests as oscillations in the loss function, which can either lead to divergence or transition into what~\cite{ahn2022understanding} terms the ``unstable convergence" (or ``edge of stability" as described in~\cite{choo2020}). At this edge of stability, although the loss function exhibits non-monotonic behavior and oscillates, optimization can still consistently decrease over a long timescale, and as a net result, gradient descent effectively continues to optimize the loss function. 

Understanding the causes of these oscillations requires insight into the shape of the loss function, particularly its sharpness, which is not readily known beforehand, especially in higher-dimensional spaces. This type of analysis exceeds the scope of the current paper. However, as further discussed in Section~\ref{sec:fluct}, in our context, oscillations primarily stem from selecting a high and multiplicatively increasing learning rate. This choice is often made to accelerate convergence and has proven beneficial in various machine learning models~\cite{smith2019super}. When the learning rate is high, the product of the sharpness and the learning rate may exceed two, potentially causing the optimization to overshoot the minimum, resulting in oscillations, unstable convergence, and eventual divergences. Several strategies can mitigate these issues, such as engineering a decreasing learning rate as optimization approaches the minimum, gradient clipping~\cite{mikolov2012statistical, pascanu2013difficulty}, cyclic learning rates as proposed in~\cite{smith2017cyclical}, or employing a decaying learning rate with cosine annealing as suggested in~\cite{loshchilov2016sgdr}. However, it's important to acknowledge that while these techniques can prevent oscillation, they may also potentially slow the optimization process.

\section{Methodology} 
\label{sec:meth}

In this work, we integrate gradient-based optimization (specifically Adam) within the variational methods framework. In practice, we consider the Gaussian basis as in the Stochastic Variational Method. The objective function to minimize is the ground state energy of the Hamiltonian in the secular equation~\eqref{eq:eigenvalue}, with respect to the nonlinear parameters in the Gaussian basis~\eqref{eq:gaussian_basis}. The gradient-based optimization enables us to devise a family of optimization strategies, broadly categorized into \emph{gradient variational methods} and \emph{hybrid methods} where random search is mixed with gradient optimization. Modifying the hyperparameters within the gradient variational methods yields a spectrum of different optimization schemes. Following a general overview of the proposed gradient variational methods, we describe their particular instances with specific hyperparameters in the subsequent sections.

\subsection{Gradient variational methods}
\label{sec:gvms}

In contrast to the baseline Stochastic Variational Method, which optimizes individual basis elements separately, in our proposed gradient variational methods, we can fix the basis size (referred to as a batch) and optimize the entire batch of basis elements using Adam. Following the optimization of this batch, two viable options are possible.

The first, which we refer to as the \emph{static gradient variational methods}, consists in iteratively adding and optimizing batches of basis functions of predefined sizes while committing to the optimized values of the previous batches. The iterative nature of this method is reminiscent of the Stochastic Variational Method iterative addition and optimization of one basis function. The static gradient variational methods are outlined in Algorithm~\ref{alg:1gvms} in Appendix~\ref{app:algs}. The second option removes the constraint of committing the previous batches. It adds the flexibility of optimizing all the previous batches simultaneously with the new proposed batch. The approach is outlined in Algorithm~\ref{alg:gvms2} in Appendix~\ref{app:algs}; we call them \emph{dynamic gradient variational methods}.
   
Similar to the Stochastic Variational Method, various stopping criteria can be proposed to halt the proposed gradient variational methods (for instance, CPU time, the difference between the energies computed in the last batches~\eqref{eq:stop_svm}, etc.). The primary goal of this paper is to compare the performances of various methods; therefore, we usually conclude the algorithm once a specified CPU runtime is reached. 

As in the Stochastic Variational Method, singularities are not eliminated in gradient variational methods. To address the ill conditions within these methods, we observe that a slight modification of the calculated gradient allows us to bypass singularities without significantly jeopardizing the optimization process. We chose to modify the gradient by generating a new one from a normal distribution with a mean value equal to the previously calculated gradient and a standard deviation of $\sigma = 0.1$.

In gradient variational methods the sizes of the batches are predetermined, and we must consider them as hyperparameters. Changing the sizes of the batches leads to different instances of the gradient variational methods. In the present paper, we evaluate and test three instances of the gradient variational methods by selecting specific values for the batches' size. 

\emph{The Baseline Gradient Variational Method (GVM)} is the simplest gradient variational method, achieved by selecting to optimize one fixed batch of $K$ basis elements. The pseudocode is outlined in Algorithm~\ref{alg:GVM} in Appendix~\ref{app:algs}.

Within the family of static gradient variational methods, the simplest involves setting each batch size equal to one. Following our chosen terminology, we refer to this approach as the \emph{Static Gradient Variational Method (Static-GVM)}, detailed in Algorithm~\ref{alg:FGVM} in Appendix~\ref{app:algs}. This method closely resembles the baseline Stochastic Variational Method, wherein each basis element is optimized individually as they are proposed, with the only distinction that the optimization is performed by Adam rather than stochastically. This approach introduces a new hyperparameter, the fixed value of the gradient optimization iterations.

Also within the dynamic gradient variational methods, we opt for testing the simplest where the batches' size is set to one. The same arguments applied to the previous method are relevant here as well. First, we decided to designate it the \emph{Dynamic Gradient Variational Method (Dynamic-GVM)}. Second, the number of optimization iterations must be predetermined and regarded as a hyperparameter (details in Algorithm~\ref{alg:DGVM} in Appendix~\ref{app:algs}).

\subsection{Hybrid method SVM-GVM}
\label{sec:hybrid} 

This method combines the baseline Stochastic Variational Method with the baseline Gradient Variational Method. Initially, we employ the Stochastic Variational Method for a specified number of basis elements (considered as a warm-up procedure for the basis set). Subsequently, we optimize the entire obtained set as in the Gradient Variational Method with Adam. This method is detailed in Algorithm \ref{alg:SVM-GVM} in Appendix~\ref{app:algs}.

Similar to the Gradient Variational Method (GVM), the hyperparameter in this method is the basis size, and among the various stopping criteria, the primary choice is the CPU runtime.

\section{Performance evaluation and analysis}
\label{sec:perfor}

\subsection{Exeprimental setup}

\subsubsection{The physical systems}

To evaluate the convergence and performance of the methods detailed in the previous Section~\ref{sec:meth}, we focus on the ground state energy of few-body systems composed of identical Bosons, particularly clusters of $\ce{He}$ up to six-body. These systems have been extensively studied and are well-replicated with potentials of Gaussian shape, as indicated in previous works~\cite{bazak:2020_Phys.Rev.A, gattobigio:2011_Phys.Rev.A, recchia2022gaussian, recchia2022subleading}. In temperature units, the energy scale associated to Helium atom is $\hbar^2/(ma_0^2) = \SI{43.28131}{K}$,  where $a_0$ is the Bohr radius, $\hbar$ the reduced Planck’s constant and $m$ the Helium atom mass. For the dimer $\ce{He_2}$, we use a two-body Gaussian potential
    \begin{equation}
        \label{eq:2b_gauss_pot_ch} 
        V(r) = V_0 \neper^{-( r/r_0)^2},
    \end{equation}
where $r$ indicates the interparticle distance, the parameters $V_0 = \SI{-1.227171}{K}$ and $r_0 = \SI{10.03019}{a_0}$ are chosen in order to reproduce the two-body observables. For systems involving three, four, five, and six-body, we consider the following two- and three-body Gaussian-shaped potentials
    \begin{align}
        & \mbox{ two-body potential: } V(r)= V_0\neper^{-(r/r_0)^2} + V_1 \left(\frac{r}{r_0} \right)^2 \neper^{-(r/r_0)^2} \label{eq:2Bnlo}\\
        & \mbox{ three-body potential: } W(\rho) = W_0 \neper^{-(\rho/\rho_0)^2} +
            W_1 \left(\frac{\rho}{\rho_0}\right)^2 \neper^{-(\rho/\rho_0)^2},
        \label{eq:3Bnlo}
    \end{align}
here, $\rho^2$ represents the hyperradius, defined as $\rho^2 = r_{12}^2 + r_{13}^2 + r_{23}^2$. The potential's parameters are specified in Table~\ref{tab:param_pot}, selected to align with three-body observables and match those obtained from the widely utilized Helium–Helium interaction model, namely, the LM2M2 potential~\cite{aziz:1991_J.Chem.Phys.}.
    
    \begin{table}[htbp]
    \centering
    \begin{tabular}{c c c c c c}
    \hline\noalign{\smallskip}
    $r_0$(a$_0$)    & $V_0$(K)   & $V_1$(K) & $W_0$(K)  & $W_1$(K) &$\rho_0$(a$_0$) \\
    \hline
    13.5            & -0.9840318 & 0.3201161 & 0.7815301 & -0.4 & 14.0 \\
    \hline\hline \noalign{\smallskip}
    \end{tabular}
    \caption{Parameters of the potential~\eqref{eq:2Bnlo} and~\eqref{eq:3Bnlo}.}
    \label{tab:param_pot}
    \end{table}    

\subsubsection{The framework and the computational resources}
\label{sec:fram_cpu}

Deep learning practitioners widely use the Adam optimizer, which has various high-performing and well-established implementations in different frameworks. We use Python within the PyTorch framework (version 2.1.0) for our implementations. It is worth noting that the Stochastic Variational Method could have been implemented with different and potentially more performing libraries. However, we opted to implement it consistently using the PyTorch framework for a fair comparison.

All the code was executed on the Intel(R) Xeon(R) CPU E5-2620 v4 @ 2.10GHz. Despite having multiple cores available on the same machine, we ran the codes on a single CPU for benchmarking purposes to accurately record the CPU runtime. 

\subsection{Overall performance evaluation}
\label{sec:first_outc}

In this section, we compare the overall performances of the baseline Stochastic Variational Method (SVM), the baseline Gradient Variational Method (GVM) and the hybrid method SVM-GVM. More details about the other methods and the chosen hyperparameters are illustrated in the next Section~\ref{sec:det_disc}.

In Table~\ref{tab:overview}, we compare the average energy, the standard deviation, and the lowest and highest energy values obtained by the original baseline Stochastic Variational Method with the baseline Gradient Variational Method and the hybrid SVM-GVM across all $n$-body systems under investigation. The stopping criteria for collecting data in this table is a given CPU runtime.

\begin{table}[htbp]
\centering
\resizebox{\textwidth}{!}{
\begin{tabular}{c|c|cccc}
\hline\hline\noalign{\smallskip}
$n$-body           & Methods  & Av. energy (K)  & Std. dv. (K) & Max. energy (K)  & Min. energy(K)\\ \hline
\multirow{2}{*}{2} & SVM      & $\SI{-3.726876e-3}{}$ & $\SI{7.29e-3}{}$  & $\SI{-1.291354e-3}{}$ & $\SI{-2.55869e-2}{}$  \\ 
                   & GVM      & $\SI{-1.293587e-3}{}$ & $\SI{2.38e-7}{}$  & $\SI{-1.226184e-3}{}$ & $\SI{-1.302777e-3}{}$ \\ 
                   & \bf SVM-GVM & $\mathbf{\SI{-1.302786e-3}{}}$ & $\mathbf{\SI{4.89e-9}{}}$  & $\mathbf{\SI{-1.302777e-3}{}}$ & $\mathbf{\SI{-1.302794e-3}{}}$  \\ 
                   \hline
\multirow{2}{*}{3} & \bf SVM      & \bf -0.1263994 & $\mathbf{\SI{1.10e-7}{}}$ & \bf -0.1263991 &  -0.1263995 \\
                   & GVM & -0.1263990 & $\SI{1.25e-7}{}$ & -0.1263953 &  -0.1263995 \\
                   & SVM-GVM & -0.1263992 & $\SI{5.00e-7}{}$ & -0.1263981 &  -0.1263995 \\ 
                   \hline
\multirow{2}{*}{4} & SVM       & -0.5665653 & $\SI{3.10e-5}{}$ & -0.566511 & -0.566605 \\
                   & GVM       & -0.5666512 & $\SI{3.69e-5}{}$ & -0.566571 & -0.566703 \\
                   & \bf SVM-GVM & \bf -0.5667353 & $\mathbf{\SI{4.56e-6}{}}$ & \bf -0.566727 & \bf -0.566741  \\ 
                   \hline
\multirow{2}{*}{5} & SVM       & -1.345126 & $\SI{1.64e-4}{}$  & -1.344697 & -1.345298  \\
                   & GVM       & -1.345787 & $\SI{3.90e-4}{}$  & -1.344813 & -1.346191 \\
                   & \bf SVM-GVM & \bf -1.346138 & $\mathbf{\SI{8.03e-05}{}}$ & \bf -1.346035 & \bf -1.346227 \\ 
                   \hline
\multirow{2}{*}{6} & SVM       & -2.484672 &  $\SI{1.92e-03}{}$ & -2.481843 & -2.487954 \\
                   & GVM       & -2.504021 &  $\SI{8.51e-04}{}$ & -2.502001 & -2.504860 \\
                   & \bf SVM-GVM & \bf -2.504888 &  $\mathbf{\SI{4.06e-4}{}}$  & \bf -2.503834 & \bf -2.505333  \\
\hline\hline\noalign{\smallskip}
\end{tabular}
}
\caption{Comparison of performance between the original Stochastic Variational Method (SVM), the baseline GVM, and the hybrid SVM-GVM for 2, 3, 4, 5, and 6-body systems. Each method was executed ten times and terminated upon reaching a specified CPU runtime limit.}
\label{tab:overview}
\end{table}

An initial observation from Table~\ref{tab:overview} indicates that the original Stochastic Variational Method consistently fails to predict the energy value for a two-body system. The variance is excessively high, and the energy values achieved are physically meaningless across multiple runs. As further explored in Section~\ref{sec:2b_sing}, this system presents significant challenges, and it is prone to produce several singularities, thereby rendering unreliable calculations in various experiments. The proposed gradient-based methods mitigate this issue by yielding more stable and reliable results.

In the case of a three-body system, where the occurrence of ill conditions is less probable, Table~\ref{tab:overview} suggests that the original Stochastic Variational Method outperforms the gradient variational methods proposed. The average and standard deviation values are marginally superior to those of the hybrid SVM-GVM method. In this case, stochastic optimization remains a viable option compared to gradient optimization. However, as further elucidated in Section~\ref{sec:fluct}, the average energy obtained at the end of the optimization procedure does not represent the lowest average energy value achieved during the optimization process. As mentioned in Section~\ref{sec:gd_optim}, during the optimization, the gradient optimization may result in energy oscillations and a non-optimal energy value. For systems comprising more than three bodies, it is safe to state that the proposed Gradient Variational Method and the hybrid SVM-GVM surpass the original Stochastic Variational Method. Both energy values and standard deviations demonstrate improvement.

\subsection{Detailed performance evaluation}
\label{sec:det_disc}

\subsubsection{Hyperparameters and performance evaluation in the two-body system}
\label{sec:2b}

For a two-body system, accurate experimental data exist~\cite{grisenti2000determination, zeller2016imaging}, and its spectrum is well-known. Table~\ref{tab:hyper_all_2b} details the selected hyperparameters for the methods investigated (the CPU time for stopping, the number of trials in the Stochastic Variational Method, the number of basis elements, the number of iterations for the gradient variational methods and the learning rate with the eventual scheduler chosen for Adam). It is important to note that the hyperparameters in this table have been fine-tuned to achieve good performance across all methods. Notably, for the Adam optimizer, we opted for a learning rate value greater than one with a linear multiplicative scheduler (the other hyperparameters are fixed as proposed in the original paper~\cite{kingma2014adam}). This choice is made to speed up the search for the minimum, particularly as we approach it. As anticipated in Section~\ref{sec:gd_optim}, using a dynamic and higher learning rate value has benefitted diverse machine learning applications~\cite{smith2019super}.

   \begin{table}[htbp]
    \centering
    \resizebox{\textwidth}{!}{
    \begin{tabular}{l| c c c c c c}
    \hline\hline\noalign{\smallskip}
    Methods & Max CPU time (s) & $\#$ Trials & $\#$ Bases & $\#$ Optim. iter. & LR & Linear scheduler \\
    \hline 
    SVM   & 20      &   100             & Time stop  & Not app.  & Not app. & Not app.  \\
    GVM   & 20      &   Not app.        &      10    & Time stop & 3.0      & 1.003  \\
    Static-GVM& 20   & Not app.          & Time stop  & 15/basis  & 1.2      & 1.01  \\
    Dynamic-GVM &20 &   Not app.        &  Time stop & 100/basis & 1.2      & 1.003 \\
    SVM-GVM & 20    &   100             &       10   & Time stop & 1.2      & 1.003 \\
    \hline\hline\noalign{\smallskip}
    \end{tabular}
    }
    \caption{Hyperparameters in the proposed methods for a two-body system.}
    \label{tab:hyper_all_2b}
    \end{table}

\begin{figure}[htbp]
    \centering
        \includegraphics[width=.8\linewidth]{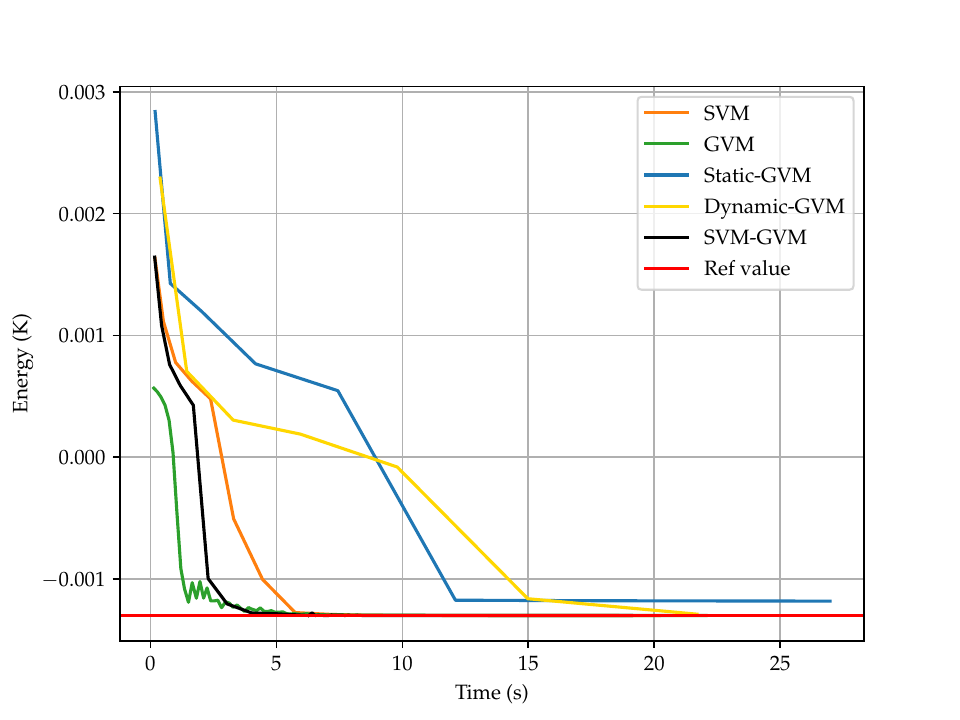}
        \caption{The energy of the best run of the proposed methods with the CPU time and the reference value for a two-body system.
        }
        \label{fig:median_all_2b}       
    \end{figure}

Some relevant outputs of the methods are detailed in Table~\ref{tab:outputs_all_2b} (the average number of basis in the Stochastic Variational Method, the average number of iterations in the gradient variational methods, the average ground state energy, and the minimum and maximum value of the energies obtained in ten different runs). Figure~\ref{fig:median_all_2b} illustrates the energy of the run returning the lowest value over ten runs by all methods.

\begin{table}[htbp]
    \centering
    \resizebox{\textwidth}{!}{
    \begin{tabular}{l|c c c c c}
    \hline\hline\noalign{\smallskip}
    Methods & Av. bases & Av. optim. iter. & Av. last energy (K) & Std. dv. (K) & Success rate \\
    \hline
    \bf SVM &      15   &       Not app.   & $\mathbf{\SI{-1.303022e-3}{}}$ & $\mathbf{\SI{1.03e-7}{}}$ & $7\%$ \\
    GVM  &    Not app.   &       170       &    $\SI{-1.293587e-3}{}$ & $\SI{2.38e-5}{}$ & $100\%$ \\
    Static-GVM &      6  &       Not app.   &    $\SI{-0.356871e-3}{}$ & $\SI{7.24e-4}{}$ & $100\%$ \\
    Dynamic-GVM &      5 &       Not app.  &    $\SI{-1.279790e-3}{}$ & $\SI{1.95e-5}{}$ & $100\%$ \\
    SVM-GVM & Not app.   &       180       &    $\SI{-1.302786e-3}{}$ & $\SI{4.88e-9}{}$ & $100\%$ \\
    \hline\hline\noalign{\smallskip}
    \end{tabular}
    }
    \caption{Some relevant outputs in the proposed methods for the chosen two-body system.}
    \label{tab:outputs_all_2b}
\end{table}

The primary conclusion drawn from Figure~\ref{fig:median_all_2b} and Table~\ref{tab:outputs_all_2b} is that the Static-GVM essentially fails to converge, or it takes an impractical amount of time to approach the ground energy value. The random search for parameters is a significant part of this method: each new basis is randomly proposed into the set and then optimized by Adam. Given that the gradient calculation is computationally expensive compared to the random search, only a limited number of optimization iterations can be set in this method (15 for each basis, as reported in Table~\ref{tab:hyper_all_2b}). The number of gradient optimization iterations is significantly smaller than the other gradient methods and the number of trials in the Stochastic Variational Method. The gradient optimization slowly converges with each new basis, necessitating more basis elements to reach the minimum. This, in turn, increases the likelihood of encountering singularities and instabilities in calculating the eigenvalues. For these reasons, Static-GVM is not included from the subsequent experiments involving more than two-body.

By examining Figure~\ref{fig:median_all_2b} and Table~\ref{tab:outputs_all_2b}, another observation regards the method Dynamic-GVM. Despite its convergence, it appears slower than the other proposed methods. This behavior is indeed expected. Each basis optimization occurs multiple times. The multiple optimization of the same basis inevitably slows the overall optimization procedure. However, as seen in Table~\ref{tab:outputs_all_2b}, the method approaches the global minimum with only five bases. Beyond being a high-performing technique, the Dynamic-GVM method is a promising tool for gaining insights about the minimum Hilbert space needed to describe a given few-body system. In the next sections, we no longer consider this method for experiments involving more than two bodies. We focus on the performance of the baseline Gradient Variational Method (GVM) and the hybrid SVM-GVM compared to the baseline Stochastic Variational Method.

We can infer a final important conclusion from the third and fourth columns of Table~\ref{tab:outputs_all_2b}. The original Stochastic Variational Method outperforms the gradient optimization techniques for the two-body system. As anticipated, the random search in a small parameter space, such as the one in the two-body system, is very fast, while calculating gradients becomes a bottleneck for the other gradient-based methods. However, it is important to underline that we had to run this algorithm multiple times to collect valuable runs due to the high occurrence of singularities in the Stochastic Variational Method (as better detailed in the next Section~\ref{sec:2b_sing}). The success rate of each method is indicated in the last column of Table~\ref{tab:outputs_all_2b}. We define a run as unsuccessful if the number of singularities exceeds the number of trials or if the calculation of the ground state eigenvalue yields meaningless results, such as an energy value lower than the actual ground state energy. As shown in this table, the success rate with the Stochastic Variational Method is only 7$\%$, implying that approximately 140 runs of the algorithm are required to obtain ten valuable results. Despite the Stochastic Variational Method being faster, the need to run it multiple times to achieve successful and useful results significantly slows down its overall performance. The gradient-based methods do not face the same challenges. Singularities are rare occurrences, and each technique succeeds in every performed run. This, in turn, makes these methods more stable and still a valuable alternative to the baseline Stochastic Variational Method, even in the smaller parameter space of the two-body systems.

\subsubsection{Hyperparameters and performance evaluation in the three-body system}
\label{sec:3b}

Similar to the two-body system, the three-body molecule of $\ce{He_3}$ has been extensively examined from a theoretical perspective~\cite{roudnev2000investigation, barletta2001variational, braaten2006universality}, and experimentally, it has been demonstrated to exhibit an Efimov-like spectrum~\cite{kunitski2015observation}. However, in contrast to the two-body scenario, the parameter space becomes more complex, with three nonlinear parameters in the Gaussian basis~\eqref{eq:gaussian_basis}. This increased complexity reduces the likelihood of encountering singularities. In all tests, singularities were never encountered with more than two bodies. Since our current focus is evaluating the methods' performance, we opt to stop them once a specified CPU time is reached. Similar to the fine-tuning process employed for the two-body system, the hyperparameters for the three-body have been chosen to ensure optimal performance, particularly the high and linear multiplicative learning rate (as indicated in column 5 and six of Table~\ref{tab:hyper_all_3b}.
    
    \begin{table}[htbp]
    \centering
    \resizebox{\textwidth}{!}{
    \begin{tabular}{l| c c c c c c}
    \hline\hline\noalign{\smallskip}
    Methods & Max CPU time (s) & $\#$ Trials & $\#$ Bases & $\#$ Optim. iter. & LR & Linear scheduler \\
    \hline 
    SVM   & 1500      &   100             & Time stop  & Not app.  & Not app. & Not app.  \\
    GVM  & 1500      &   Not app.        &      40    & Time stop & 3.0      & 1.003  \\
    SVM-GVM & 1500   &   100             &       40   & Time stop & 1.2      & 1.003 \\
    \hline\hline\noalign{\smallskip}
    \end{tabular}
    }
    \caption{Hyperparameters in the proposed methods for a three-body system.}
    \label{tab:hyper_all_3b}
    \end{table}

\begin{figure}[t!]
    \centering
        \includegraphics[width=.8\linewidth]{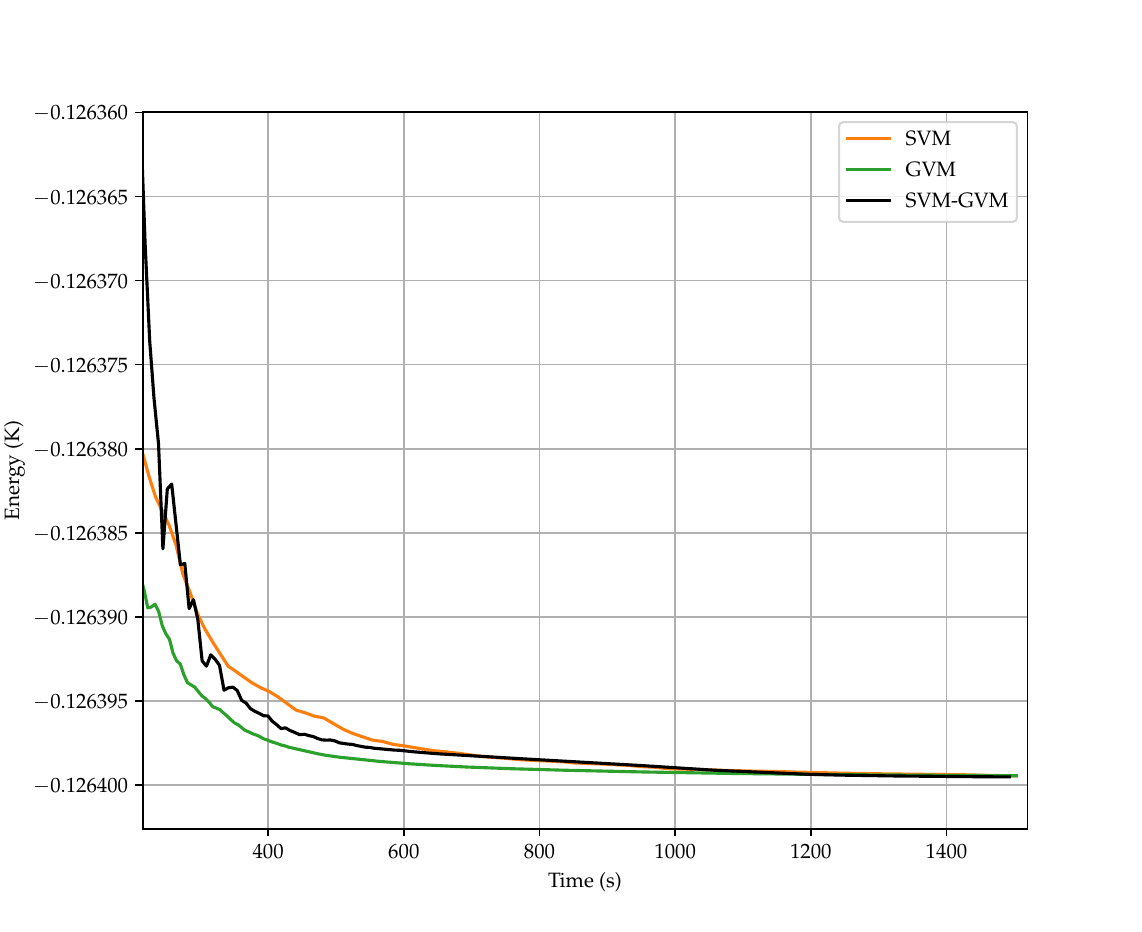}
        \caption{The energy of the best run of the proposed methods with the CPU time for a three-body system. The tail of the graph has been magnified for better clarity.
        }
        \label{fig:median_all_3b}       
    \end{figure}

Table~\ref{tab:outputs_all_3b} shows relevant outputs for the chosen three-body system. Figure~\ref{fig:median_all_3b} depicts the energy along the CPU time of the run returning the lowest value obtained over ten runs.

   \begin{table}[htbp]
    \centering
    \begin{tabular}{l| c c c c c}
    \hline\hline\noalign{\smallskip}
    Methods & Av. bases & Av. optim. iter. & Av. last energy (K) & Std. dv. (K) \\
    \hline
    \bf SVM &      123      &       Not app.        &\bf -0.1263994 & $\mathbf{\SI{1.10e-7}{}}$  \\
    GVM  &    Not app.   &       270             &    -0.1263990 & $\SI{1.25e-6}{}$   \\
    SVM-GVM & Not app.   &       200             &    -0.1263992 & $\SI{5.00e-7}{}$  \\
    \hline\hline\noalign{\smallskip}
    \end{tabular}
    \caption{Some relevant outputs in the proposed methods for the chosen three-body system.}
    \label{tab:outputs_all_3b}
    \end{table}

When considering the average energy achieved at the end of the optimization process, the third and fourth column in Table~\ref{tab:outputs_all_3b} leads to the same conclusion as in the two-body system: the Stochastic Variational Method continues to outperform gradient optimization techniques in the three-body system as well, both in term of average energy and variance. Nevertheless, the average energy values obtained, particularly in the Gradient Variational Method and SVM-GVM, are very close to those obtained by the Stochastic Variational Method, with only $0.003\permil$ and $0.0002\permil$ of separation, respectively. Moreover, as elucidated in Section \ref{sec:3b}, it is essential to highlight that these energy values represent the mean of the energy obtained at the end of the optimization procedure. As mentioned in Section~\ref{sec:gd_optim}, gradient optimization methods may result in energy oscillation and unstable convergence; therefore, the final energy is not always the minimum energy. This, according to the variational principle, cannot happen in the Stochastic Variational Method since by adding a new basis element, the new energy is always lower than the previous one. The phenomenon of energy oscillation is further explored in Section~\ref{sec:fluct}.

\subsubsection{Hyperparameters and performance evaluation with four-, five-body and six-body}
\label{sec:456b}

Very seldom do experimental data exist for more than three-body. The potential we use in four-, five- and six-body is the same as in Equation~\eqref{eq:2Bnlo} and~\eqref{eq:3Bnlo}, and the parameters are the same as in Table~\ref{tab:param_pot}. This potential has been shown to accurately approach the same ground energy values' obtained by the Helium–Helium LM2M2 interaction~\cite{recchia2022subleading}. 

As chosen in the previous sections, we mainly focus on the methods' performance. We stop the algorithms once a specified CPU time is reached. The hyperparameters of the methods are presented in Table~\ref{tab:hyper_all_45b}. Unlike the two- and three-body systems, there is no need for the same fine-tuning of the reported hyperparameters: with four-, five- and six-body systems, we observe fast convergence of energy with less sensitivity of the choice of hyperparameters. 

    \begin{table}[htbp]
    \centering
    \resizebox{\textwidth}{!}{
    \begin{tabular}{l| c c c c c c}
    \hline\hline\noalign{\smallskip}
    \multicolumn{7}{c}{{\bf 4-body}} \\
    \hline\noalign{\smallskip}
    Methods & Max CPU time (s) & $\#$ Trials & $\#$ Bases & $\#$ Optim. iter. & LR & Linear scheduler \\
    \hline
    SVM   &       2500       &   100             & Time stop  & Not app.  & Not app. & Not app.  \\
    GVM  &       2500       &   Not app.        &  25        & Time stop &   0.8    & 1.0       \\
    SVM-GVM &    2500       &   100             &  25        & Time stop &   0.8    & 1.0        \\
    \hline\hline\noalign{\smallskip}
    \multicolumn{7}{c}{{\bf 5-body}} \\
    \hline\noalign{\smallskip}
    Methods & Max CPU time (s) & $\#$ Trials & $\#$ Bases & $\#$ Optim. iter. & LR & Linear scheduler \\
    \hline 
    SVM   &       10000     &   100             & Time stop  & Not app.  & Not app. & Not app.  \\
    GVM  &       10000     &   Not app.        &  25        & Time stop & 0.8      & 1.0       \\
    SVM-GVM &    10000     &   100             &  25        & Time stop & 0.8       & 1.0       \\
    \hline\hline\noalign{\smallskip}
    \multicolumn{7}{c}{{\bf 6-body}} \\
    \hline\noalign{\smallskip}
    Methods & Max CPU time (s) & $\#$ Trials & $\#$ Bases & $\#$ Optim. iter. & LR & Linear scheduler \\
    \hline 
    SVM   &       19000     &   100             & Time stop  & Not app.  & Not app. & Not app.  \\
    GVM   &       19000     &   Not app.        &  25        & Time stop & 0.8      & 1.0       \\
    SVM-GVM  &    19000     &   100             &  25        & Time stop & 0.8       & 1.0       \\
    \hline\hline\noalign{\smallskip}
    \end{tabular}
    }
    \caption{Hyperparameters in the SVM, GVM, and SVM-GVM for four-, five- and six-body systems.}
    \label{tab:hyper_all_45b}
    \end{table}

\begin{figure}[htbp]
    \centering
        \includegraphics[width=.8\linewidth]{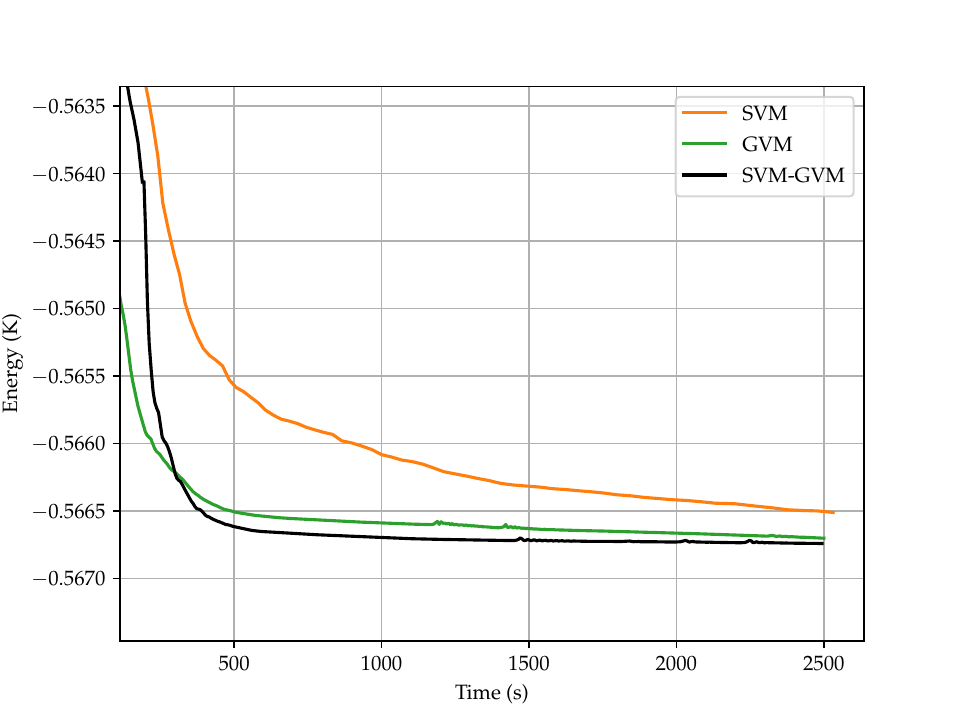}
        \caption{The energy of the best run of the SVM, GVM, and SVM-GVM with the CPU time in a four-body system. The tail of the graphs has been magnified for better clarity.
        }
        \label{fig:median_4B_125}       
    \end{figure}

\begin{figure}[htbp]
    \centering
        \includegraphics[width=.8\linewidth]{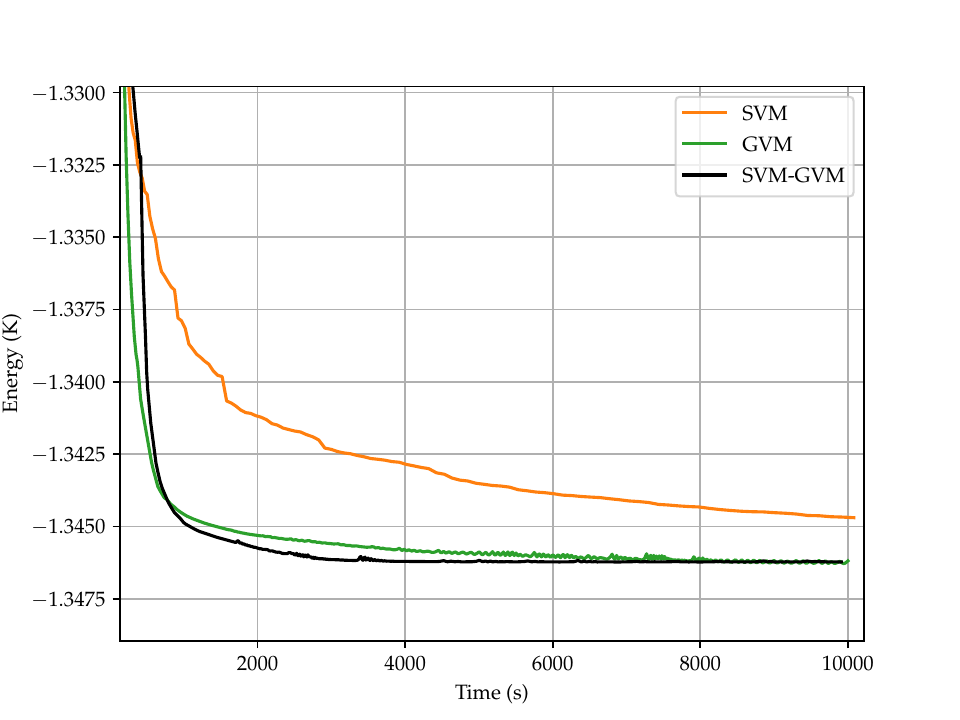}
        \caption{The energy of the best run of the SVM, GVM, and SVM-GVM with the CPU time in a five-body system. The tail of the graph has been magnified for better clarity.
        }
        \label{fig:median_5B_125}       
    \end{figure}

\begin{figure}[htbp]
    \centering
        \includegraphics[width=.8\linewidth]{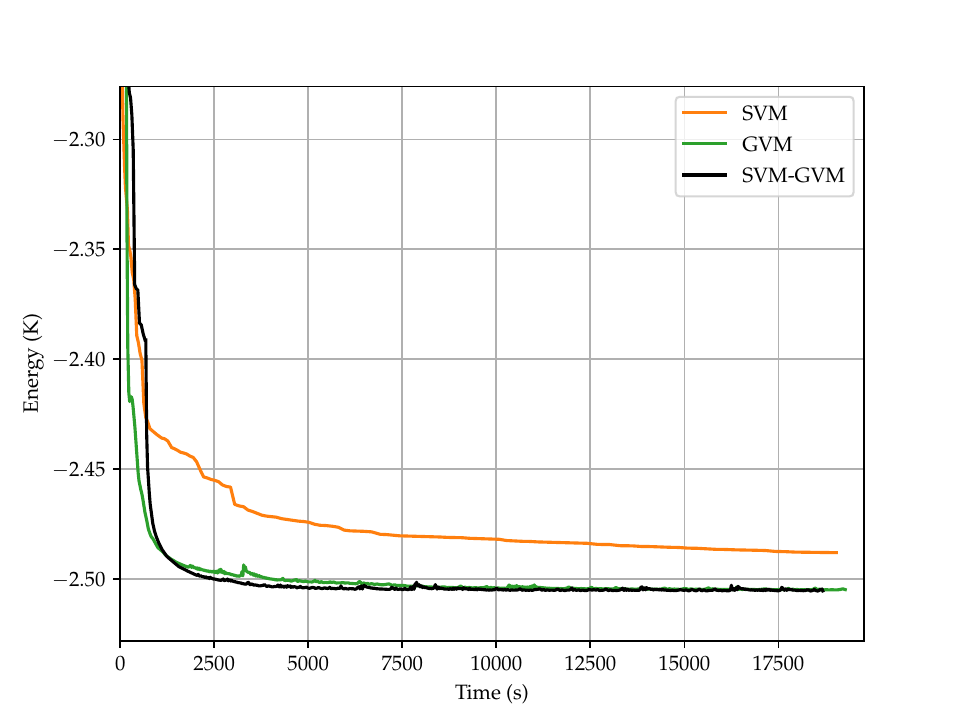}
        \caption{The energy of the best run of the SVM, GVM and SVM-GVM with the CPU time in a six-body system. The tail of the graph has been magnified for better clarity.
        }
        \label{fig:median_6B_15}       
    \end{figure}

Table~\ref{tab:outputs_all_45b} shows some relevant outputs, and Figures~\ref{fig:median_4B_125}, \ref{fig:median_5B_125} and~\ref{fig:median_6B_15} illustrate the energy of the run resulting in the lowest value of SVM, GVM, and SVM-GVM over ten runs in four-, five- and six-body systems respectively.

   \begin{table}[htbp]
    \centering
    \begin{tabular}{l| c c c c}
    \hline\hline\noalign{\smallskip}
    \multicolumn{5}{c}{{\bf 4-body}} \\
    \hline\noalign{\smallskip}
    Methods & Average bases & Average optim. iter. & Av. energy (K) & Std. dv. (K) \\
    \hline
    SVM   &      93       &       Not app.        &    -0.5665653   & $\SI{3.10e-5}{}$ \\
    GVM  &    Not app.   &       390             &    -0.5666512   & $\SI{3.69e-5}{}$ \\
    \bf SVM-GVM & Not app.&       360            & \bf -0.5667353  & $\mathbf{\SI{4.56e-6}{}}$ \\
    \hline\hline\noalign{\smallskip}
    \multicolumn{5}{c}{{\bf 5-body}} \\
    \hline\noalign{\smallskip}
    Methods & Average bases & Average optim. iter. & Av. last energy (K) & Std. dv. (K)  \\
    \hline
    SVM   &      125      &       Not app.        &    -1.345126   & $\SI{1.64e-4}{}$ \\
    GVM  &    Not app.   &       700             &    -1.345787   & $\SI{3.90e-4}{}$ \\
    \bf SVM-GVM & Not app.   &       650         &\bf -1.346138   & $\mathbf{\SI{8.03e-5}{}}$ \\
    \hline\hline\noalign{\smallskip}
    \multicolumn{5}{c}{{\bf 6-body}} \\
    \hline\noalign{\smallskip}
    Methods & Average bases & Average optim. iter. & Av. last energy (K) & Std. dv. (K) \\
    \hline
    SVM   &      125      &       Not app.        &  -2.484672   & $\SI{1.92e-3}{}$ \\
    GVM   & Not app.      &       850             &  -2.504021  & $\SI{8.51e-4}{}$ \\
    \bf SVM-GVM & Not app.   &       750         & \bf -2.504888  & $\mathbf{\SI{4.06e-4}{}}$ \\
    \hline\hline\noalign{\smallskip}
    \end{tabular}
    \caption{Some relevant outputs in the SVM, GVM, and SVM-GVM in four-, five- and six-body systems.}
    \label{tab:outputs_all_45b}
    \end{table}

    As evident in Table~\ref{tab:outputs_all_45b} and Figures~\ref{fig:median_4B_125},~\ref{fig:median_5B_125} and~\ref{fig:median_6B_15} the gradient variational methods outperform the original Stochastic Variational Method in four-, five-body and six-body systems in each conducted runs. Notably, the hybrid SVM-GVM method is the most efficient, converging faster with a smaller variance. From Table \ref{tab:outputs_all_45b}, we observe that the improvement in average energy is approximately $0.3\permil$ in four-body, $0.8\permil$ in five-body and $8\permil$ in six-body. \emph{As expected, the gradient methods exhibit enhanced performance compared to random search as the dimension of the search space increases.}

\subsection{Singularities in the Stochastic Variational Methods and challenging systems}
\label{sec:2b_sing}

The study of the two-body system merits particular attention for several reasons. Firstly, only one nonlinear parameter must be chosen in the Gaussian basis~\eqref{eq:gaussian_basis}, making the random choice in the Stochastic Variational Method efficient. However, the space of nonlinear parameters is very small, and as mentioned in Section~\ref{sec:bgSVM}, the occurrence of singularities is more likely to happen in the random choice. From a physics perspective, this system is very shallow, and its ground state energy is close to zero, indicating weak binding. This property makes it computationally challenging to find the ground state eigenvalue of the Hamiltonian, further increasing the likelihood of encountering singularities. For all these reasons, the two-body system is the ideal candidate to assess the stability of the proposed gradient variational methods with the Stochastic Variational Method.

\begin{figure}
    \centering
        \includegraphics[width=.8\linewidth]{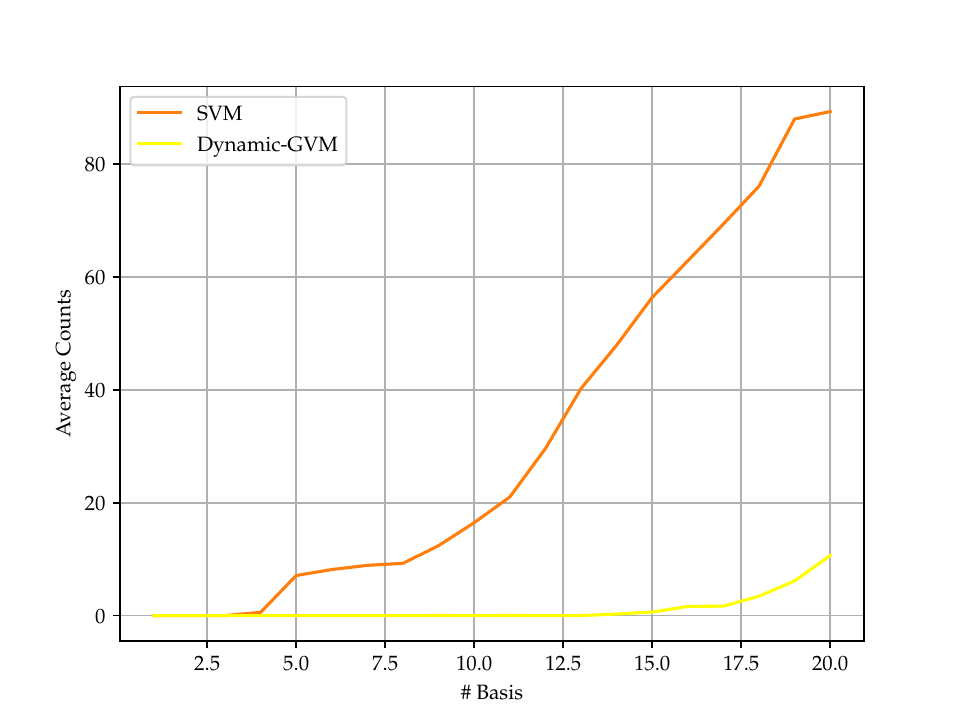}
        \caption{Average counts of singularities occurrences over 100 runs of the SVM and the Dyanamic-GVM.}
    \label{fig:sing_2b_svm_dfsgo}       
    \end{figure}

In this section, we compare the amount of singularities occurrences between the Stochastic Variational Method and the Dynamic-GVM. Both methods are stopped after reaching $K = 20$ basis elements. To ensure a fair comparison, we set the number of trials $L$ in the Stochastic Variational Method to be the same as the number of gradient optimization iterations $L$ in the Dynamic-GVM, with both set to one hundred. We repeated the two methods one hundred times and recorded the count of singularity occurrences on each basis. Figure~\ref{fig:sing_2b_svm_dfsgo} illustrates the average count of singularities at each basis.

Two essential considerations arise from Figure~\ref{fig:sing_2b_svm_dfsgo}. First, we see that the number of singularities with the Dynamic-GVM is less than that of the original Stochastic Variational Method. Second, we observe that the amount of singularities increases as the basis size and then as we approach the ground state energy value. It is worth noting that, in the Dynamic-GVM, Adam attempts to find the global minimum supported by a given number of bases. As we approach the minimum, we can expect a higher chance of encountering singularities. Nevertheless, the gradient method remains reasonably stable even in this scenario, further reinforcing the assertion about its improved stability compared to the Stochastic Variational Method.

Theoretical investigation into why gradient variational methods are less prone to singularities is beyond the scope of the current paper. However, a potential explanation lies in the search method employed. Randomly jumping into a space affected by singularities is ``riskier" compared to smoothly exploring it by following the steepest slope of the energy landscape. 

\subsection{Oscillations in gradient variational methods} 
\label{sec:fluct}
  
As anticipated in Section~\ref{sec:gd_optim}, the value of the learning rate plays a crucial role in the neighborhood of a minimum. Specifically, if the product of the loss function's sharpness (the highest eigenvalue of its Hessian) and the learning rate exceeds two, the optimization process oscillates. This can lead to either divergences or a phase known as unstable convergence (or the edge of stability), where the optimization converges non-monotonically over a long timescale~\cite{choo2020,ahn2022understanding}. In optimising the quantum few-body ground state energy via gradient variational methods, we also encounter divergences and unstable convergence instances.

\begin{figure}[t!]
    \centering
        \includegraphics[width=.8\linewidth]{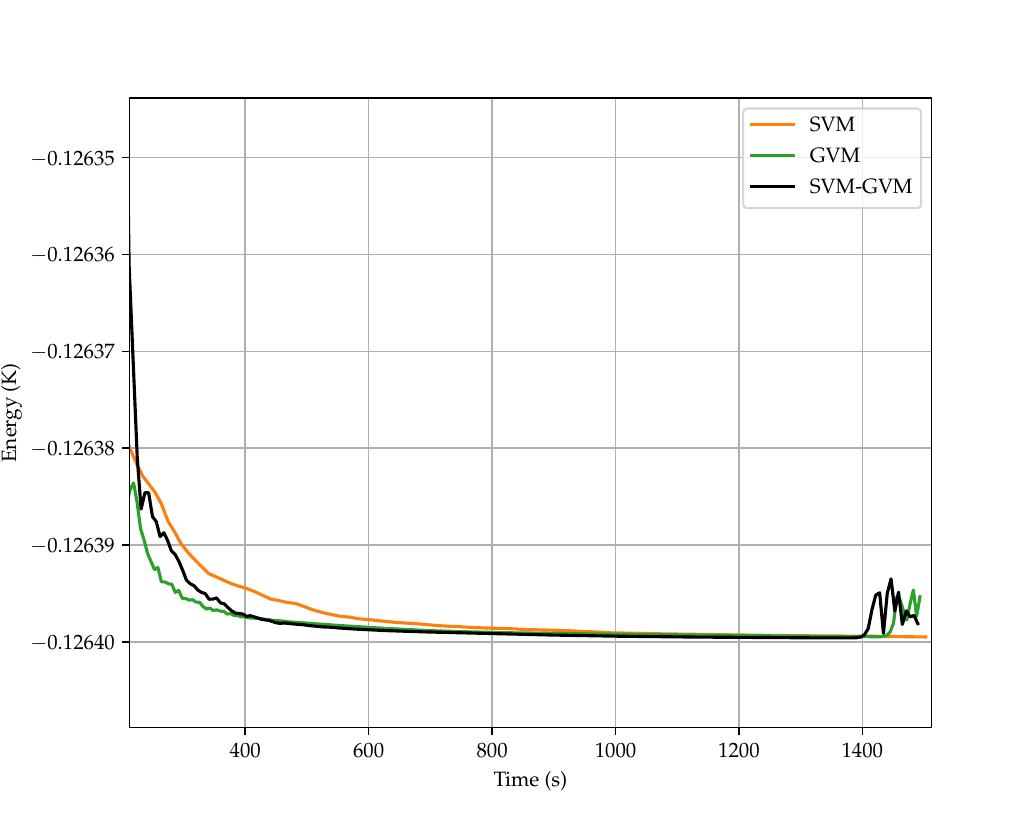}
        \caption{The energies of the methods SVM, GVM, and SVM-GVM in the worst over ten runs, with the corresponding CPU time and the reference value for a three-body system.}
        \label{fig:only_max_3B_125}       
    \end{figure}

We encounter divergences when the gradient approaches the region of the minimum near the end of the optimization procedure. In this scenario, if CPU runtime is used as the stopping criterion, the optimization continues even after reaching the optimum. As noted in Section~\ref{sec:3b}, Table~\ref{tab:outputs_all_3b} suggests that, when considering the average energy at the end of the optimization, the Stochastic Variational Method outperforms the gradient variational methods GVM and SVM-GVM. However, the final energy obtained using gradient variational methods is not always the minimum energy reached during the optimization. Figure~\ref{fig:only_max_3B_125} shows the energies obtained in the worst runs, specifically those converging to the highest energy values, for the baseline Stochastic Variational Method and the gradient variational methods GVM and SVM-GVM.
    
As observed in Figure \ref{fig:only_max_3B_125}, these gradient variational methods outperform the original Stochastic Variational Method, and approaching the global minimum, they tend to overshoot it, resulting in the energy's oscillation. In our case, this phenomenon is primarily attributed to the high and multiplicative increasing learning rate (as seen with the linear multiplicative learning rate chosen for the three-body system in Table~\ref{tab:hyper_all_3b}), which can result in a high value of the sharpness multiplied by the learning rate, causing the oscillations and divergences. To mitigate the increasing learning rate different workaround have been proposed in the literature (for the sake of examples, we can cite cyclic learning rates~\cite{smith2017cyclical} or a decaying learning rate with cosine annealing~\cite{loshchilov2016sgdr}, clipping the gradient~\cite{mikolov2012statistical, pascanu2013difficulty}). However, it is important to note that while these techniques may mitigate the oscillations, they potentially slow the optimization process. For the current analysis, in the first column of Table~\ref{tab:outputs_min_3b}, we present the average minimum energy obtained by each method. The values closely align with those obtained using the baseline Stochastic Variational Method. Specifically, the SVM-GVM method slightly outperforms the original Stochastic Variational Method, with a comparable variance. Regardless of the oscillation around the global minimum, we can state that the hybrid SVM-GVM method slightly outperforms the baseline Stochastic Variational Method in a three-body system as well.

   \begin{table}[htbp]
    \centering
    \begin{tabular}{l| c c}
    \hline\hline\noalign{\smallskip}
    Methods     & Av. min. energy (K) & Std. dv. (K) \\
    \hline
    SVM         & -0.1263994        & $\mathbf{\SI{1.10e-7}{}}$  \\
    GVM         &    -0.1263994     & $\SI{1.37e-7}{}$ \\
    \bf{SVM-GVM}   & \bf -0.1263995 &  $\SI{1.18e-7}{}$  \\
    \hline\hline\noalign{\smallskip}
    \end{tabular}
    \caption{The average and the standard deviation of the minimum energy obtained throughout the optimization process in the proposed methods for the chosen three-body system.}
    \label{tab:outputs_min_3b}
    \end{table}

\begin{figure}[t!]
    \centering
        \includegraphics[width=1\linewidth]{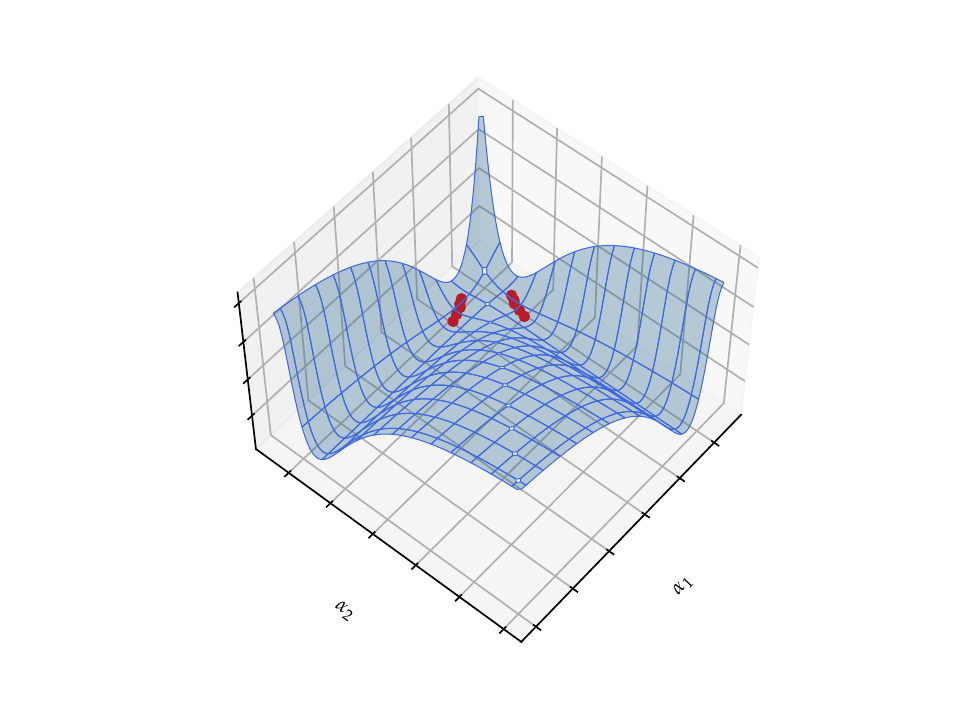}
        \caption{The landscape of the energy in case of two nonlinear parameters with two-body and two bases. The red dots indicate the point where all the surrounding points in the grid are higher (the global minima).}
        \label{fig:2b-2B}       
    \end{figure}

Besides the divergences observed near the global minimum, the Figures~\ref{fig:median_all_2b}, \ref{fig:median_all_3b}, \ref{fig:median_4B_125}, \ref{fig:median_5B_125}, and \ref{fig:median_6B_15} also show energy oscillations during the optimization process. However, these oscillations do not compromise the optimization, and although the energy decreases non-monotonically, it converges over a long timescale. Understanding the primary cause of these oscillations requires knowledge of the energy landscape, particularly whether it is affected by different local minima and their sharpness.  Exploring the entire space of nonlinear parameters to identify local minima and evaluate their sharpness is computationally challenging. For our current analysis, we only investigate the presence of local minima within a grid in selected cases where the space is small enough to make this assessment feasible. Specifically, for the two-body system with one and two basis elements, we did not encounter local minima within the chosen grid, apart from a few cases, due to numerical error. Figure~\ref{fig:2b-2B} depicts the energy landscape concerning the two nonlinear parameters for a two-body system with two bases. The landscape shown in the figure appears smooth, with two distinct symmetric wide valleys (the global minima represented by the red dots in the figure). This simple analysis suggests that the oscillations may be attributed again to the high learning rate of the gradient optimization step rather than the presence of multiple local minima or sharp minima. Nevertheless, we cannot conclusively assert that the space of nonlinear parameters is free of multiple local minima or sharp minima. In higher-dimensional space, the deepness of the minima may differ in different directions, and verifying the presence of local minima and evaluating their sharpness becomes computationally infeasible.

\section{Conclusion}

The Stochastic Variational Method is a stochastic search method leveraging the variational principle. It is an efficient, accurate, state-of-the-art method for solving various few-body quantum systems. It is, however, challenged by singularities when the number of particles is small and by the size of the search space when the number of particles increases.

We investigate the opportunity to replace stochastic search with gradient-based optimization. We propose and compare three gradient-based methods (named gradient variational methods) and a hybrid method that mixes the Stochastic Variational Method with a gradient variational method. We evaluate the strengths and weaknesses of these algorithms for the computation of the ground state energy of two-, three-, four-, five and six-body systems containing identical bosons, specifically clusters of helium $\ce{He}$, which can be effectively described using Gaussian-shaped potentials.

Our results for the two-body system show that the gradient variational methods exhibit improved stability, resulting in better performance. Indeed, fewer singularities occur during the optimization process, suggesting a potential advantage in incorporating gradient-based techniques into studying challenging physics systems. 

The Stochastic Variational Method outperforms the proposed gradient variational methods for the computation of the energy of a three-body system. However, when employing gradient variational methods, we observe occasional oscillations and divergences in the energy value, particularly when approaching the minima. This highlights the importance of finely tuning the learning rate in the gradient descent algorithm to prevent its increasing. Nevertheless, upon excluding the oscillations from our analysis, the gradient variational methods, particularly the hybrid SVM-GVM method, show competitive performance that surpasses the results obtained by the Stochastic Variational Method.
 
In exploring four-, five- and six-body systems, the tested gradient variational methods consistently outperform the Stochastic Variational Method, demonstrating their efficiency in higher-dimensional spaces.

The outcomes of this study suggest that while the original Stochastic Variational Method remains powerful, gradient variational methods offer promising alternatives. These techniques show potential in addressing the computational challenges encountered in the Stochastic Variational Method, paving the way for more efficient solutions to the quantum few-body problem. The observed success of gradient variational methods in higher-dimensional systems implies their scalability, encouraging further exploration in even more complex physics scenarios.

For the sake of completeness, we should acknowledge the potential further enhancement in both the Stochastic Variational Method and our proposed gradient variational methods. For example, Stochastic Variational Methods lend themselves to parallelization, where multiple trials can be conducted concurrently, with the minimum selected from among these trials, while gradient variational methods can be refined by incorporating analytical expressions of gradients whenever possible.

We are now exploring alternative basis sets with nonlinear parameters, such as wavelets as well as neural network ansaetze. The interplay between random and gradient-based search strategies could contribute to refining these methods for broader use in quantum few-body physics. 

\section*{Acknowledgements}
\footnotesize
This research is supported by Singapore Ministry of Education, grant MoE-T1251RES2302, and by the National Research Foundation, Prime Minister’s Office, Singapore, under its Campus for Research Excellence and Technological Enterprise (CREATE) programme as part of the programme Descartes. D. Basu acknowledges the support of the ANR JCJC project REPUBLIC (ANR-22-CE23-0003-01), and the PEPR project FOUNDRY (ANR23-PEIA-0003).
\normalsize


\appendix

\section{List of Algorithms}
\label{app:algs}

In this Appendix, we report the pseudocodes of the algorithms described and tested in the paper:

\begin{enumerate}
    \item Stochastic Variational Method (SVM)
    \item Static gradient variational methods
    \item Dynamic gradient variational methods
    \item Baseline Gradient Variational Method (GVM)
    \item Static Gradient Variational Method (Static-GVM)
    \item Dynamic Gradient Variational Method (Dynamic-GVM)
\end{enumerate}

\begin{algorithm}[H]
    \caption{Stochastic Variational Method (SVM)}
    \label{alg:or_SVM}
    \SetKw{KwGoTo}{go to}

    \KwData{$K$ = basis size, $L$ = number of trials, $H$ = Hamiltonian}
    
    \For{basis $ k = 1$ \KwTo K}{
        
        \For{number of trials $l = 1$ \KwTo L}{
                      
            Pick as the number of pair $\frac{N(N-1)}{2}$ random nonlinear parameters $\alpha^{(k)}_{ab} \sim U(0, \alpha_{max}]$ \label{random_param}
        
            Make up the basis element $\psi^{(l)}_k$
        
            \For{each index $j \geq i = 1$ \KwTo k}{
                Being $\psi_k =\psi^{(l)}_k$
                
                Calculate $\mathcal{S}^{(l)}_{ij} = \braket{\psi_i | \psi_j}$ and   $\mathcal{S}^{(l)}_{ji} = \mathcal{S}^{(l)}_{ij}$
            
                Calculate $\mathcal{H}^{(l)}_{ij} = \braket{\psi_i | H| \psi_j}$ and $\mathcal{H}^{(l)}_{ji} = \mathcal{H}^{(l)}_{ij}$ 
            }
            
            Solve the eigenvalue problem $\mathcal{H}^{(l)}\mathcal{C} = E^{(l)} \mathcal{S}^{(l)}\mathcal{C}$
            
            \If{$\mathcal{S}^{(l)}$ is singular}{
                    \KwGoTo \ref{random_param}\;
                 }
            }
            Set $l = \underset{l}{\mathrm{argmin}} [E_{1}^{(l)}]$
            
            Set $\psi_k = \psi_k^{(l)}$
            
            \KwRet $E_1 = E_1^{(l)}$
    }

\end{algorithm}

    \begin{algorithm}[H]
    \caption{Static gradient variational methods}
    \label{alg:1gvms}
        \SetKw{KwGoTo}{go to}
        \SetKwProg{Fn}{Function}{:}{}
        
        \KwData{$\bm K = K_1, K_2, \dots $ = batches' size, $H$ = Hamiltonian}

        \SetKwFunction{FillOverlapHamiltonian}{FillOverlapHamiltonian}
        
        Epoch $m = 1$

        Initialize $K_{0} = 0$ 
        
        \Repeat{A given condition is met}{
            
            Pick at random a batch of $\frac{N(N-1)}{2} \times K_m$ nonlinear parameters $\bm \alpha_{K_m}$ \label{random_gvms}

            From $\bm \alpha_{K_m}$ make up the batch  $\bm A_{K_m} = \{\psi_k\}_{k=K_{m-1}+1, \dots, K_{m-1}+K_m}$ 
            
            $\mathcal{S}, \mathcal{H} $ = \FillOverlapHamiltonian{$\bm A_{K_m}$}
           
            Solve the eigenvalue problem $\mathcal{H}\mathcal{C} = E \mathcal{S}\mathcal{C}$
            
            \If{$\mathcal{S}$ is singular}{
                \KwGoTo \ref{random_gvms}\;
            }
            
            \Repeat{A given condition is met}{
                Being the lowest eigenvalue the loss function $E_1(\bm \alpha_{K_m})$

                Calculate the gradient of the loss function $\nabla E_1(\bm \alpha_{K_m})$
                                    
                Update $\bm \alpha_{K_m}$ based on Adam optimizer \label{adam_gvms}
                
                From the new $\bm \alpha_{K_m}$ make up the batch  $\bm A_{K_m} = \{\psi_k\}_{k=K_{m-1}+1, \dots, K_{m-1}+K_m}$

                $\mathcal{S}, \mathcal{H} $ = \FillOverlapHamiltonian{$\bm A_{K_m}$}
           
                Solve the eigenvalue problem $\mathcal{H}\mathcal{C} = E \mathcal{S}\mathcal{C}$
            
                \If{$\mathcal{S}$ is singular}{
                    Modify the gradient extracting it from a normal distribution $\nabla E_1(\bm \alpha_{K_m}) \sim N(\nabla E_1(\bm \alpha_{K_m}), 0.1)$
                    
                    \KwGoTo \ref{adam_gvms}\;
                }
            }
            
            Next epoch $m = m + 1$
        }
                  
        \KwRet $E_1$

        \Fn{\FillOverlapHamiltonian{$\bm A_{K_m}$}}{
                
            \For{each index $j \geq i = K_{m-1}+1$ \KwTo $K_{m-1}+K_m$}{
                
            Calculate $\mathcal{S}{ij} = \braket{\psi_i | \psi_j}$ and $\mathcal{S}_{ji} = \mathcal{S}_{ij}$
            
            Calculate $\mathcal{H}_{ij} = \braket{\psi_i | H| \psi_j}$ and $\mathcal{H}_{ji} = \mathcal{H}_{ij}$
            }

            \KwRet $\mathcal{S}, \mathcal{H}$
        }  
    \end{algorithm}

    
    \begin{algorithm}[H]
        \caption{Dynamic gradient variational methods}
        \label{alg:gvms2}
        \SetKw{KwGoTo}{go to}
        \SetKwProg{Fn}{Function}{:}{}

        \KwData{$\bm K = K_1, K_2, \dots $ = batches' size, $H$ = Hamiltonian}

        \SetKwFunction{FillOverlapHamiltonian}{FillOverlapHamiltonian}
        
        Epoch $m = 1$

        Initialize $K_{0} = 0$ 
        
        \Repeat{A given condition is met}{
            
            Pick at random a batch of $\frac{N(N-1)}{2} \times K_m$ nonlinear parameters $\bm \alpha_{K_m}$ \label{random_gvms2}

            Append to the entire set $\bm \alpha \leftarrow \bm \alpha_{K_m}$
            
            From $\bm \alpha$ make up the batch  $\bm A = \{\psi_k\}_{k=0, \dots, K_{m-1}+K_m}$ 
            
            $\mathcal{S}, \mathcal{H} $ = \FillOverlapHamiltonian{$\bm A$}
           
            Solve the eigenvalue problem $\mathcal{H}\mathcal{C} = E \mathcal{S}\mathcal{C}$
            
            \If{$\mathcal{S}$ is singular}{
                \KwGoTo \ref{random_gvms2}\;
            }
            
            \Repeat{A given condition is met}{
                Being the lowest eigenvalue the loss function $E_1(\bm \alpha)$

                Calculate the gradient of the loss function $\nabla E_1(\bm \alpha)$
                                    
                Update $\bm \alpha$ based on Adam optimizer \label{adam_gvms2}
                
                From the new $\bm \alpha$ make up the basis  $\bm A = \{\psi_k\}_{k=0, \dots, K_{m-1}+K_m}$

                $\mathcal{S}, \mathcal{H} $ = \FillOverlapHamiltonian{$\bm A$}
           
                Solve the eigenvalue problem $\mathcal{H}\mathcal{C} = E \mathcal{S}\mathcal{C}$
            
                \If{$\mathcal{S}$ is singular}{
                    Modify the gradient extracting it from a normal distribution $\nabla E_1(\bm \alpha) \sim N(\nabla E_1(\bm \alpha), 0.1)$
                    
                    \KwGoTo \ref{adam_gvms2}\;
                }
            }
            
            Next epoch $m = m + 1$
        }
                  
        \KwRet $E_1$

        \Fn{\FillOverlapHamiltonian{$\bm A$}}{
                
            \For{each index $j \geq i = 1$ \KwTo $K_{m-1}+K_m$}{
                
            Calculate $\mathcal{S}{ij} = \braket{\psi_i | \psi_j}$ and $\mathcal{S}_{ji} = \mathcal{S}_{ij}$
            
            Calculate $\mathcal{H}_{ij} = \braket{\psi_i | H| \psi_j}$ and $\mathcal{H}_{ji} = \mathcal{H}_{ij}$
            }

            \KwRet $\mathcal{S}, \mathcal{H}$
        }  
    \end{algorithm}


        \begin{algorithm}[H]
        \caption{Gradient Variational Method (GVM)}
        \label{alg:GVM}
        \SetKw{KwGoTo}{go to}
        \SetKwProg{Fn}{Function}{:}{}

        \KwData{$K$ = basis size, $H$ = Hamiltonian}

        \SetKwFunction{FillOverlapHamiltonian}{FillOverlapHamiltonian}
            
        Pick $K \times \frac{N(N-1)}{2}$ random nonlinear parameters $\bm \alpha$ \label{random_param_gvm}

        From $\bm \alpha$ make up the set of basis elements $\bm A = \{\psi_k\}_{k=1,\dots,K}$

        $\mathcal{S}, \mathcal{H} $ = \FillOverlapHamiltonian{$\bm A$}
           
        Solve the eigenvalue problem $\mathcal{H}\mathcal{C} = E \mathcal{S}\mathcal{C}$
            
        \If{$\mathcal{S}$ is singular}{
            \KwGoTo \ref{random_param_gvm}\;
            }

        \Repeat{A given condition is met}{
            Being the lowest eigenvalue the loss function $E_1(\bm \alpha)$

            Calculate the gradient of the loss function $\nabla E_1(\bm \alpha)$
                
            Update $\bm \alpha$ based on Adam optimizer \label{adam_gvm}

            From the new $\bm \alpha$ make up the basis elements $ \bm A = \{\psi_k\}_{k=1,\dots,K}$
            
            $\mathcal{S}, \mathcal{H} $ = \FillOverlapHamiltonian{$\bm A$}
                            
            Solve the eigenvalue problem $\mathcal{H}\mathcal{C} = E \mathcal{S}\mathcal{C}$

            \If{$\mathcal{S}$ is singular}{
                    Modify the gradient extracting it from a normal distribution $\nabla E_1(\bm \alpha) \sim N(\nabla E_1(\bm \alpha), 0.1)$
                        
                    \KwGoTo \ref{adam_gvm}\;
                }
            }  
        \KwRet $E_1$

        \Fn{\FillOverlapHamiltonian{$\bm A$}}{
                
        \For{each index $j \geq i = 1$ \KwTo k}{
                
            Calculate $\mathcal{S}{ij} = \braket{\psi_i | \psi_j}$ and   $\mathcal{S}_{ji} = \mathcal{S}_{ij}$
            
            Calculate $\mathcal{H}_{ij} = \braket{\psi_i | H| \psi_j}$ and $\mathcal{H}_{ji} = \mathcal{H}_{ij}$

            \KwRet $\mathcal{S}, \mathcal{H}$
            }
        }

        \end{algorithm}

        \begin{algorithm}[H]
        \caption{Static Gradient Variational Method (Static-GVM)}
        \label{alg:FGVM}
        \LinesNumbered
        \SetKw{KwGoTo}{go to}

        \KwData{$L$ = number of optimization iterations, $H$ = Hamiltonian}

        Epoch $m=1$
        
        \Repeat{Until a condition is met}{
                
            Pick as the number of pairs $\frac{N(N-1)}{2}$ random nonlinear parameters $\alpha^{(m)}_{ab}$ \label{random_param_sgvm}
        
            \For{number of step $l = 1$ \KwTo L}{

                Calculate the corresponding basis vector $\psi_m$    
                   
                \For{each index $j = 1$ \KwTo m}{
                             
                    Calculate $\mathcal{S}_{mj} = \braket{\psi_m | \psi_j}$ and $\mathcal{S}_{jm} = \mathcal{S}_{mj}$
                
                    Calculate $\mathcal{H}_{mj} = \braket{\psi_m | H| \psi_j}$ and $\mathcal{H}_{jm} = \mathcal{H}_{mj}$
                    }
                    
                Solve the eigenvalue problem $\mathcal{H}\mathcal{C} = E\mathcal{S}\mathcal{C}$

                \If{$\mathcal{S}$ is singular $\And l = 0$}{
                        \KwGoTo \ref{random_param_sgvm}\;
                    }

                \If{$\mathcal{S}$ is singular $\And l > 0$}{
                    Modify the gradient extracting it from a normal distribution $\nabla E_1(\alpha^{(m)}_{ab}) \sim N(\nabla E_1(\alpha^{(m)}_{ab}), 0.1)$
                            
                        \KwGoTo \ref{adam_sgvm}\;
                }
                    
                Being the lowest eigenvalue the loss function $E_1(\alpha^{(m)}_{ab})$

                Calculate the gradient of the loss function $\nabla E_1(\alpha^{(m)}_{ab})$

                Update $\alpha^{(m)}_{ab}$ based on Adam optimizer \label{adam_sgvm}
                
            }
                
            Increase epoch $m = m + 1$
        }
            
        \KwRet $E_1$
            
        \end{algorithm}

        \begin{algorithm}[H]
        \caption{Dynamic Gradient Variational Method (Dynamic-GVM)}
        \label{alg:DGVM}
        \LinesNumbered
        \SetKw{KwGoTo}{go to}
        \SetKwProg{Fn}{Function}{:}{}

        \KwData{$L$ = number of optimization iterations, $H$ = Hamiltonian}

        \SetKwFunction{FillOverlapHamiltonian}{FillOverlapHamiltonian}

         Epoch $m=1$
         
        \Repeat{A given condition is met}{
                
            Pick as the number of pairs $\frac{N(N-1)}{2}$ random nonlinear parameters $\alpha^{(m)}_{ab}$ \label{random_param_dgvm}

            Append to the set of nonlinear parameters $\bm \alpha \leftarrow \alpha^{(m)}_{ab}$
                
            Make up the basis element $\psi_m$

            Append to the set of bases $\bm A \leftarrow \psi_m$
    
            $\mathcal{S}, \mathcal{H} $ = \FillOverlapHamiltonian{$\bm A$}
                
            Solve the eigenvalue problem $\mathcal{H}\mathcal{C} = E \mathcal{S}\mathcal{C}$

            \If{$\mathcal{S}$ is singular}{
                \KwGoTo \ref{random_param_dgvm}\;
            }

            \For{number of optimization iterations $l = 1$ \KwTo L}{
                Being the lowest eigenvalue the loss function $E_1(\bm \alpha)$

                Calculate the gradient of the loss function $\nabla E_1(\bm \alpha)$
                                    
                Update $\bm \alpha$ based on Adam optimizer \label{adam_dgvm}

                From the new $\bm \alpha$ make up the basis set $\bm A = \{\psi_k\}_{k=1,\dots,m}$
            
                $\mathcal{S}, \mathcal{H} $ = \FillOverlapHamiltonian{$\bm A$}
            
                Solve the eigenvalue problem $\mathcal{H}\mathcal{C} = E \mathcal{S}\mathcal{C}$

                \If{$\mathcal{S}$ is singular}{
                    Modify the gradient extracting it from a normal distribution $\nabla E_1(\bm \alpha) \sim N(\nabla E_1(\bm \alpha), 0.1)$
                        
                    \KwGoTo \ref{adam_dgvm}\;
                    }
            }
            Increase epoch $m = m + 1$
        }
                  
        \KwRet $E_1$

        \Fn{\FillOverlapHamiltonian{$\bm A$}}{
                
            \For{each index $j \geq i = 0$ \KwTo m}{
                
            Calculate $\mathcal{S}{ij} = \braket{\psi_i | \psi_j}$ and   $\mathcal{S}_{ji} = \mathcal{S}_{ij}$
            
            Calculate $\mathcal{H}_{ij} = \braket{\psi_i | H| \psi_j}$ and $\mathcal{H}_{ji} = \mathcal{H}_{ij}$

            \KwRet $\mathcal{S}, \mathcal{H}$
            }
        }  
    \end{algorithm}

    \begin{algorithm}[H]
        \caption{SVM-GVM}
        \label{alg:SVM-GVM}
        \LinesNumbered
        \SetKw{KwGoTo}{go to}

        \KwData{$K$ = basis size, $H$ = Hamiltonian}

        \tcp{Warm up with Stochastic Variational Method}

        Perform the Algorithm \ref{alg:or_SVM} for $K$ basis elements
            
        Return the set of nonlinear parameters $\bm \alpha = \{\alpha^{(k)}_{ab}\}_{k=1,\dots,K}$ and the corresponding basis set $\bm A = \{\psi_k\}_{k=1,\dots,K}$
            
        \tcp{Optimization with the Gradient Variational Method} 

        Perform the Algorithm \ref{alg:GVM} with the initialized set of nonlinear parameters $\bm \alpha$ and corresponding basis set $\bm A$

    \end{algorithm}

\section{The secular equation}
\label{app:sec_eq}

Let us approximate the eigenvector $\psi_l$ with the ansatz $\phi_l$ as in~\eqref{eq:system_vecs}
\begin{equation}
\label{eq:a_eigvec}
\psi_l\approx \phi_l = \sum_{k=1}^{K} c_{kl} \psi_k.
\end{equation} 
For the $l$th eigenvalue, the functional~\eqref{eq:functional} can be written as
\begin{equation}
\label{eq:functional_Ak}
    E_l[\phi_l] = \frac{\braket{\displaystyle\sum_{k=1}^{K} c_{kl}\psi_k | H | \displaystyle\sum_{k=1}^{K} c_{kl} \psi_k}}{\braket{\displaystyle\sum_{k=1}^{K} c_{kl} \psi_k | \displaystyle\sum_{k=1}^{K} c_{kl} \psi_k}} = \frac{\displaystyle\sum_{i=1}^{K} \displaystyle\sum_{j=1}^{K} c^*_{il} c_{jl} \mathcal{H}_{ij} }{\displaystyle\sum_{i=1}^{K} \displaystyle\sum_{j=1}^{K} c^*_{il} c_{jl} \mathcal{S}_{ij}},
\end{equation}
where 
\begin{align}
\label{eq:ham_mat}
    &\mathcal{H}_{ij} = \braket{\psi_i | H | \psi_j} \\   
\label{eq:overlap_mat}
    & \mathcal{S}_{ij} = \braket{\psi_i | \psi_j},
\end{align}
are the matrix elements of the Hamiltonian and overlap matrices, respectively, in the chosen basis. According to the Ritz theorem, the parameters $c_{kl}$ in \eqref{eq:a_eigvec} can be determined by the stationary condition
\begin{equation}
\label{eq:stationary_cond_def}
    \delta E_l[\phi_l] = 0.
\end{equation}
The variation of the functional \eqref{eq:functional_Ak} with respect to the parameters $c_{kl}$ in \eqref{eq:a_eigvec}, leads to the stationary condition
\begin{equation}
\label{eq:stationary_cond}
    \frac{\partial E_l[\phi_l] }{\partial c^*_{kl}} = \frac{\displaystyle\sum_{j=1}^{K} c_{jl} ( \mathcal{H}_{kj} - E_l\mathcal{S}_{kj})}{\displaystyle\sum_{i=1}^{K} \displaystyle\sum_{j=1}^{K} c^*_{il} c_{jl} \mathcal{S}_{ij}} = 0 \implies \sum_{j=1}^{K} \mathcal{H}_{kj} c_{jl} = E_l \sum_{j=1}^{K} \mathcal{S}_{kj} c_{jl}.
\end{equation}
Introducing the eigenvector matrix $\mathcal{C}$ and the diagonal matrix of eigenvalues $E = \text{diag}(E_l)_{l=1,\dots, K}$, Equation \eqref{eq:stationary_cond} becomes
\begin{equation}
(\mathcal{HC})_{kl} = (\mathcal{SC}E)_{kl},
\end{equation}
which represents a linear system of $K$ equations in $c_{kl}$. For each eigenvalue $E_l$ this system can be solved as a generalized finite eigenvalue problem, often referred to as the secular equation~\eqref{eq:eigenvalue}
\begin{equation}
\mathcal{HC} = \mathcal{SC}E.
\end{equation}

\section{Solving the secular equation} 
\label{app:solv_sec_equ}

The primary goal is to reduce the task of solving a generalized eigenvalue problem as the one formulated in~\eqref{eq:eigenvalue} in a standard eigenvalue problem. We could achieve this purpose by using the symmetric orthogonalization of the basis. 

Instead of the old basis set $\bm A = \{\psi_k\}_{k=1,\dots, K}$ in which the matrices $\mathcal{H}{ij} = \braket{\psi_i | H | \psi_j}$ and $\mathcal{S}{ij} = \braket{\psi_i | \psi_j}$ are calculated, we could use an orthogonal basis set defined as
\begin{equation}
    \bm A' = \mathcal{S}^{-\frac{1}{2}} \bm A.
\end{equation}
The matrix $\mathcal{S}^{-\frac{1}{2}}$ is computed as follow
\begin{enumerate}
    \item We diagonalize $\mathcal{S}$ with a unitary matrix  
    \begin{equation}
        \mathcal{S}_\textup{diag} = U^\dag \mathcal{S} U.
    \end{equation}  
    \item Since the matrix $\mathcal{S}$ is symmetric positive definite, the eigenvalues of $\mathcal{S}$ are all positive. We can safely calculate the square roots of the diagonal elements in $\mathcal{S}_\textup{diag}$ that, in turn, allow us to define the symbol $\mathcal{S}_\textup{diag}^{\frac{1}{2}}$.
    \item The matrix $\mathcal{S}^{\frac{1}{2}}$ is then defined as 
    \begin{equation}
        \mathcal{S}^{\frac{1}{2}} = U \mathcal{S}_\textup{diag}^{\frac{1}{2}}  U^\dag. 
    \end{equation} 
    \item Finally the matrix $\mathcal{S}^{-\frac{1}{2}}$ is defined as 
    \begin{equation}
    \label{eq:s12}
        \mathcal{S}^{-\frac{1}{2}} = (\mathcal{S}^{\frac{1}{2}})^{-1} =   U \mathcal{S}_\textup{diag}^{-\frac{1}{2}}  U^\dag.
    \end{equation} 
\end{enumerate}

By multiplying both left side of~\eqref{eq:eigenvalue} by $\mathcal{S}^{-\frac{1}{2}}$ defined in Equation~\eqref{eq:s12} and considering that $\mathcal{S}^{-\frac{1}{2}} \mathcal{S} = \mathbb{I}$,  after some calculation we obtain
\begin{equation}
    \mathcal{S}^{-\frac{1}{2}}  \mathcal{HC} = E \mathcal{S}^{-\frac{1}{2}} \mathcal{SC} \implies  \mathcal{S}^{-\frac{1}{2}}  \mathcal{H} \mathcal{S}^{-\frac{1}{2}} \mathcal{S}^{\frac{1}{2}} C = E \mathcal{S}^{\frac{1}{2}} \mathcal{C}
\end{equation}
which can be rewritten as a standard eigenvalue problem
\begin{equation}
\label{eq:eigenvalue_std}
    \Tilde{\mathcal{H}} \Tilde{\mathcal{C}} = E \Tilde{\mathcal{C}}
\end{equation}
where
\begin{align}
    & \Tilde{\mathcal{H}} = \mathcal{S}^{-\frac{1}{2}}  \mathcal{H} \mathcal{S}^{-\frac{1}{2}} \\
    & \Tilde{\mathcal{C}} = \mathcal{S}^{\frac{1}{2}} \mathcal{C}.
\end{align}
The standard eigenvalue problem~\eqref{eq:eigenvalue_std} has the same spectrum of eigenvalues as the generalized eigenvalue problem~\eqref{eq:eigenvalue}.

\bibliographystyle{apalike}
\bibliography{bib_improve_SVM}

\end{document}